\documentclass[sn-mathphys,Numbered]{sn-jnl}

\usepackage{graphicx}%
\usepackage{multirow}%
\usepackage{amsmath,amssymb,amsfonts}%
\usepackage{amsthm}%
\usepackage{mathrsfs}%
\usepackage[title]{appendix}%
\usepackage{xcolor}%
\usepackage{textcomp}%
\usepackage{manyfoot}%
\usepackage{booktabs}%
\usepackage{algorithm}%
\usepackage{algorithmicx}%
\usepackage{algpseudocode}%
\usepackage{listings}%

\usepackage{comment}

\theoremstyle{thmstyleone}%
\theoremstyle{thmstyletwo}%

\theoremstyle{thmstylethree}%

\begin{document}

\title[Article Title]{An interpretable and transferable model for shallow landslides detachment combining spatial Poisson point processes and generalized additive models}


\author*[1]{\fnm{Giulia} \sur{Patanè}}\email{giulia.patane@polimi.it}

\author[1]{\fnm{Teresa} \sur{Bortolotti}}\email{teresa.bortolotti@polimi.it}

\author[2]{\fnm{Vasil} \sur{Yordanov}}\email{vasil.yordanov@polimi.it}

\author[2]{\fnm{Ludovico Giorgio Aldo} \sur{Biagi}}\email{ludovico.biagi@polimi.it}

\author[2]{\fnm{Maria Antonia} \sur{Brovelli}}\email{maria.brovelli@polimi.it}

\author[3]{\fnm{Xuan Quang} \sur{Truong}}\email{txquang@hunre.edu.vn}

\author[1]{\fnm{Simone} \sur{Vantini}}\email{simone.vantini@polimi.it}

\affil[1]{\orgdiv{MOX, Department of Mathematics}, \orgname{Politecnico di Milano}, \orgaddress{\country{Italy}}}

\affil[2]{\orgdiv{Department of Civil and Environmental Engineering}, \orgname{Politecnico di Milano}, \orgaddress{\country{Italy}}}

\affil[3]{\orgdiv{Information Technology Faculty}, \orgname{Hanoi University of Natural Resources and Environment}, \orgaddress{\country{Vietnam}}}


\abstract{Less than 10 meters deep, shallow landslides are rapidly moving and strongly dangerous slides. In the present work, the probabilistic distribution of the landslide detachment points within a valley is modelled as a spatial Poisson point process, whose intensity depends on geophysical predictors according to a generalized additive model. Modelling the intensity with a generalized additive model jointly allows to obtain good predictive performance and to preserve the interpretability of the effects of the geophysical predictors on the intensity of the process. We propose a novel workflow, based on Random Forests, to select the geophysical predictors entering the model for the intensity. In this context, the statistically significant effects are interpreted as activating or stabilizing factors for landslide detachment. In order to guarantee the transferability of the resulting model, training, validation, and test of the algorithm are performed on mutually disjoint valleys in the Alps of Lombardy (Italy). Finally, the uncertainty around the estimated intensity of the process is quantified 
via semiparametric bootstrap.}




\maketitle

\section{Introduction}\label{sec1}
Shallow landslides, characterized by depth not exceeding 10 meters and often leading to rapid and destructive movements, are extensively studied for their great danger. With the aim of minimizing the loss of lives and properties, landslide susceptibility maps for a given geographical area are widely used to inform about the risk of landslides occurring in the area. Landslide susceptibility maps are raster images associating each pixel to the probability that the pixel is crossed by a landslide \cite{loche2022}. A lot of work has been dedicated to the construction of covariate-based models for landslide susceptibility maps of shallow landslides (e.g., \cite{land_sus_validated, article_SL1, article_SL2}).
A typical aspect of state-of-the-art landslide susceptibility mapping concerns the fact that no distinction is made between the pixels related to the detachment of a landslide and the pixels that are solely interested by its passage. Our methodological proposal stems from noticing that if a pixel has only been a point of passage for landslides, the covariates associated to the pixel should not be considered as activating or stabilising factors of the landslide, since the latter was likely generated in a point at higher altitude. The typical models used for landslide susceptibility originate in the field of machine learning. In \cite{SVM_SL}, the authors employ Support Vector Machine for the classification of the pixels as at susceptible to landslides or not. In identifying the optimal support vector machine for classification, covariate selection is performed. In \cite{RF_SL} Random Forests are used for classification and simultaneously to quantify the importance of the covariates in the classification task. In \cite{Adaboost_SL_validno} decision trees and Adaboost are combined for the first time in landslide susceptibility modeling, while in \cite{CNN_SL} the authors employ Convolutional Neural Networks to gain a high predictive power. In \cite{biagi2021}, a review of the different machine learning-based approaches for calculating landslide susceptibility is proposed.
All of these methods are marked by their significant flexibility and their capability to capture the complexity of the covariates effects; however, they rarely offer an easy interpretation of the relationship between the geophysical covariates and susceptibility. Consequently, they do not allow to associate a high susceptibility area with the specific value of one or more geophysical covariates, hindering the identification of targeted intervention strategies to reduce the vulnerability of the area to shallow landslides. Another aspect that today is rarely sought is transferability in new areas. In the context of landslide risk assessment, few works validate the estimated models in unseen geographical areas. In \cite{validation_paper}, for example, the authors propose temporal, spatial and random partition of the data in order to use part of it for training and part of it to validate the estimated model. However, the partition is always performed on a unique connected valley. In \cite{land_sus_validated}, in \cite{land_sus_validated2} and in \cite{Adaboost_SL_validno} the dataset is randomly partitioned for training and validation, but the data always refer to the same area. In \cite{SVM_SL}, the training and testing are performed in the same valley, respectively in the year 1992 and in 1999. We point out that by selecting the covariates in the same valley considered for training, the obtained models are likely to become valley-specific and thus of limited applicability. Moreover, from a statistical perspective, testing on the same valley used for training might result in a strong underestimation of the actual prediction error. t With the attempt of bringing an alternative perspective to landslides risk assessment, we propose a new conceptual framework for shallow landslide modeling. We model the probability that a pixel is the detachment point of a landslide, which is hereafter referred to as \textit{crown}. We call \textit{landslide detachment map} the map that associates each pixel to the probability of being a landslide detachment point of a shallow landslide. The landslide detachment map deeply differs from the landslide susceptibility map, in that the former associates each pixel with the probability that it is the point of detachment of a landslide, the latter with the probability that the pixel is crossed by a landslide. In the first case the pixel is \textit{active} in the landslide generation, while in the second case it is \textit{passive} to it. The reason why we propose the use and estimation of landslide detachment maps instead of a landslide susceptibility maps is that we expect that the relationship between the geophysical characteristics of a pixel with the probability that it is a landslide detachment point is more meaningful than the one with the probability that it is crossed by a landslide. Formally, we model the shallow landslide \textit{crowns}, i.e. the highest parts of the main scarp \cite{land_glossary}, as a realization of a spatial Poisson point process \cite{baddeley06} with covariates, a geostatistical model that allows to simulate the spatial distribution of the crowns based on geophysical predictors. The intensity of the process, which determines the infinitesimal probability of each location to be detachment point of a landslide, is assumed to depend on the geophysical characteristics of the locations following a generalized additive model. The landslide detachment map coincides, according to the estimated model, with the intensity map of the process. By integrating the intensity over a specific area, one gets the expected number of crowns in the area. Crowns, being the positions where the landslides start their sliding towards valley, are the most informative points of landslides of shallow type. Moreover the significant effects of the predictors on our model can be seen as generating or stabilizing factors, since they influence the risk of crown presence, i.e. the risk of landslide generation. The effects of the predictors are highly interpretable by using generalized additive models for the estimation of the intensity of the crowns spatial process. One of our main objective is the transferability of our models to new areas, where there is no inventory of landslides, and for this reason we devote great attention to the validation (phase in which model selection is performed), testing and uncertainty estimation of predictions. In particular, three disjoint valleys are used, respectively, to train, validate and test the models. We propose a nonparametric estimate of the uncertainty of the predictions based on bootstrapping. We also illustrate and use a method to guide the selection of covariates, through the use of Random Forests \cite{RF}. The potential to model the landslide crowns spatial pattern is to be able, in future developments, to integrate it with a numerical simulation of the dynamics of landslides (from the crown to the valley), so as to obtain a complete modeling of the landslide phenomena of shallow type. Today, the research of models that are able to simulate the dynamics of landslides is strongly active \cite{GATTI2024112798, GATTI2023105362, quecedo04, pastor17, pastor21, GATTI2024128525}. Notice that by combining a model that simulate the landslide detachment points with one that simulate the dynamics of the landslide, it is also possible to provide much more reliable susceptibility maps than the ones which are currently built using state of the art approach. The paper is structured as follows. In Section \ref{data}, the dataset is illustrated, and the preprocessing on it is explained. In Section \ref{method} the Poisson Spatial Point Process with GAM intensity is defined, together with the bootstrap-based estimation method for the uncertainty quantification of predictions. In Section \ref{results} the results on our training, validation and test valleys are shown and discussed. The conclusions are discussed in Section \ref{conclusion}.

\section{Data exploration and preprocessing}
\label{data}

\subsection{Data}
\begin{figure}[H]
     \centering
     \includegraphics[scale=0.4]{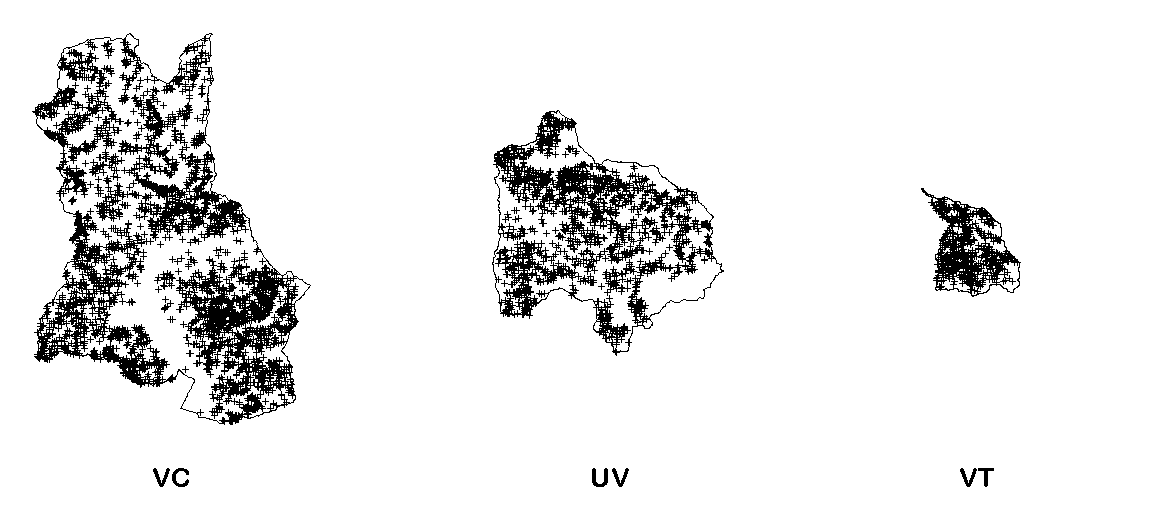}
     \caption{Landslide crowns inventory: data are relative to three separate valleys in the province of Sondrio, Italy, i.e. (from left to right) Val Chiavenna, Upper Valtellina and Val Tartano. The points represent the positions of the crowns.}
     \label{lci}
 \end{figure}

Data are relative to three separate valleys in the province of Sondrio, Italy, i.e. Val Chiavenna, Upper Valtellina and Val Tartano (see Figure \ref{lci}). For each of them, we are provided with the historical inventory of the position of the shallow landslides crowns and a selection of geophysical information, collected on a pixel-by-pixel basis into a 5-meter resolution raster data. The geophysical quantities considered are: digital terrain model (\textit{DTM}), \textit{slope}, \textit{eastness}, \textit{northness}, topographic wetness index (\textit{TWI}), normalized difference vegetation index (\textit{NDVI}), profile curvature (\textit{PRC}), plan curvature (\textit{PLC}), distance from roads, distance from rivers, distance from faults and land use/cover (\textit{DUSAF}). 
The first eight covariates are continuous, while the distances are ordinal covariates with 5 levels and \textit{DUSAF} is categorical with 11 classes. 
The \textit{DTM} indicates the elevation above the sea, in meters. \textit{Northness} and \textit{eastness} are, respectively, the cosine and the sine of the topographic aspect, i.e. the direction that the slope is facing. The \textit{TWI} refers to the potential ability of the ground to absorb water based on the topography. Typically, the value of \textit{TWI} indicators have range $(-3,30)$ \cite{twi_report}.
The \textit{NDVI} is a dimensionless index representing the density of green on an area of land if the value is positive, and the density of water if the value is negative. We point out that presence of water can indicate both the presence of lake or river and the presence of snow. Generally the values of \textit{NDVI} in presence of snow are, in absolute value, lower than those of lake/river \cite{ndvi_report}. The \textit{PRC} and the \textit{PLC} are, respectively, the second derivative of the vertical and horizontal section of the mountain side. Henceforth, the term \textit{landslide} will imply \textit{shallow landslide}.

\subsection{Preprocessing}
\begin{figure}[H]
    \centering
    \includegraphics[scale=0.4]{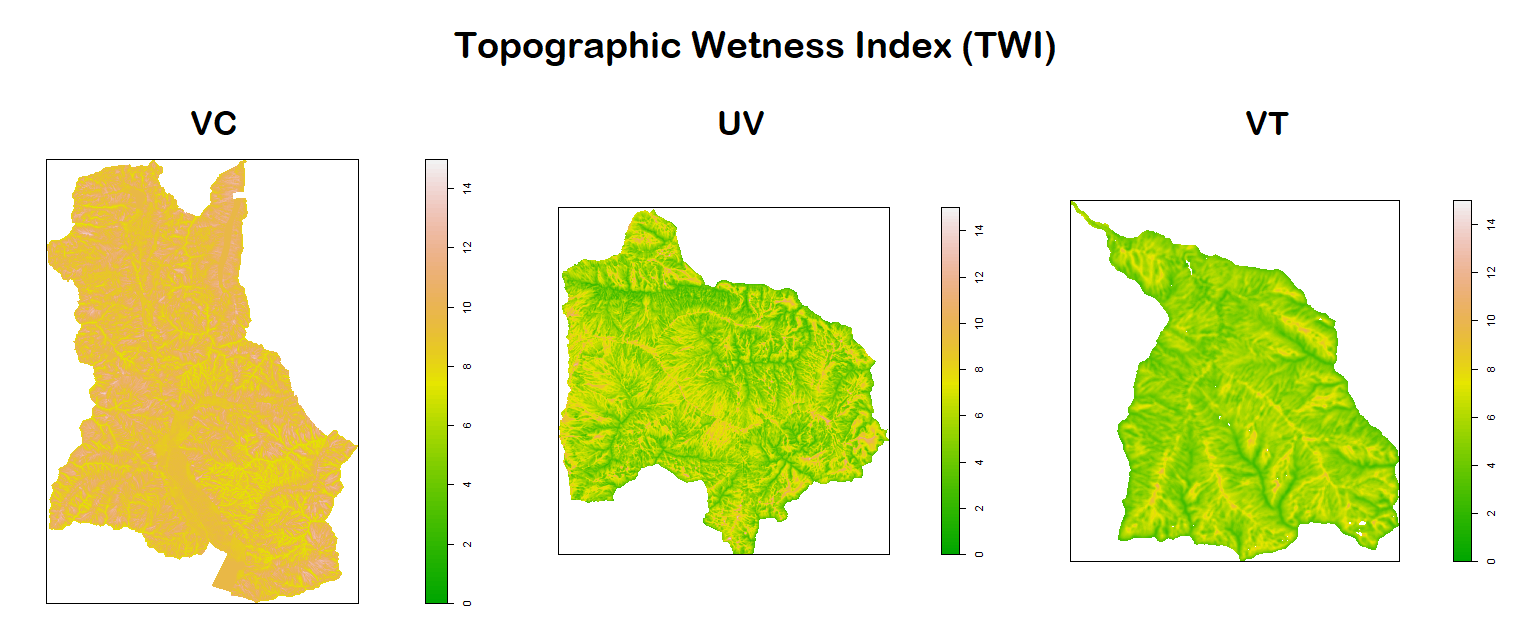}
    \caption{Variable \textit{TWI} in the three valleys: Val Chiavenna, Upper Valtellina and Val Tartano. Val Chiavenna consistently displays a higher range of $TWI$ compared to the other two valleys.}
    \label{twi_compare}
\end{figure}

As the crown point is a pointwise approximation of a detachment area, it is convenient to associate the point with a weighted average of the geophysical covariates of the pixels in the area around it. More specifically, the averaging is performed by applying a Gaussian filter to the following continuous covariates: \textit{DTM}, \textit{slope}, \textit{northness}, \textit{eastness}, \textit{TWI}, \textit{NDVI}. The Gaussian filter is set with standard deviation equal to 100 $m$ and radius equal to 10 $m$. Recall that the radius represents the distance beyond which the contribute to the weighted average is zero. A second type of preprocessing was performed on \textit{TWI}, and is motivated by the fact that \textit{TWI} spans different ranges in the three considered valleys. In particular, Val Chiavenna consistently displays higher vallues of $TWI$ with respect to the other two valleys (see Figure \ref{twi_compare}). In order to prevent model instability issues due to the unmatching ranges of $TWI$ in the three valleys, we propose to use a new covariate, \textit{$TWI_b$}, synthetizing the information provided by the original \textit{TWI}. More specifically, the new variable is defined as $TWI_b = 0$ if $TWI \leq 9$, and $TWI_b = 1$ otherwise. The value $TWI=9$ is a threshold value for the topographic wetness index, beyond which the wetness is significant from a geophysical point of view \cite{twi_report}.
\noindent
A third preprocessing is done on plan curvature and profile curvature, and grounds on the fact that it is the sign of these continuous covariates, rather than their absolute value, which most impacts the likelihood of occurrence of shallow landslides [REF]. For this reason, and in a model economy perspective, the covariates are converted into cathegorical covariates with three levels: \textit{negative}, \textit{zero}, \textit{positive}. The \textit{DUSAF} covariate is categorical with 11 classes. Notice that, knowing \textit{NDVI}, we are able to know if a zone presents snow, water or vegetation and to quantify their presence; hence, thanks to the covariate \textit{NDVI}, it is possible to discriminate among the following seven categories: inland waters, inland wetlands, sparse or absent vegetation areas, evolving vegetation areas, non-agricoltural green, permanent lawns and forests. Notice that all these classes represent a non-urbanized area, so we merged them into a unique class: \textit{Natural}. The remaining four classes are: urban zone, production sites, mines/landfills/work sites and arable land. We merged them into the class \textit{Anthropic}. The result is a binary \textit{DUSAF}, where we do not lose information about the amount of vegetation/water/snow in the natural lands, since we know it from the \textit{NDVI}, and we merge the less prevalent areas (the anthropic ones) into a more general category. In the Anthropic class, the \textit{NDVI} is typically around 0 since there is neither a frequent nor remarkable presence of snow, water or vegetation.

\section{Methodology}
\label{method}
In this section, we propose the methodology to model the patterns of crowns within a given geographical area. We model the crown distribution as a non-homogeneous point process. Unlike homogeneous point processes where points are distributed uniformly, in a non-homogeneous process, the spatial density of points varies across the space. By integrating the geophysical covariates into our modeling framework, we seek to capture and to understand the complex interactions between the environmental characteristics and the crowns generation.
\subsection{Spatial Poisson Point Processes with GAM intensity}
A spatial point process \cite{baddeley06} is a random pattern of points in a $d$-dimensional space, with $d\geq2$. Spatial point processes are useful statistical models for the analysis of observed patterns of points, where the points represent the locations of some object of study. Any spatial point process is fully characterized by its random counting measure
\begin{equation}
    N(A)= \text{number of points falling in A},
\end{equation}
where $A$ is a bounded closed set of $\mathbb{R}^d$. In many applications, a useful summary statistic of point processes is given by the expected value of $N$, which we refer to as \textit{intensity measure}
\begin{equation}
    \nu(A) =\mathbb{E}[N(A)].
\end{equation}
When the derivative of $\nu$ exists, it is convenient in terms of interpretation to directly estimate the density of the intensity measure, i.e. the function $\lambda: \mathbb{R}^d \mapsto \mathbb{R}^+$ such that
\begin{equation}
\nu(A)=\displaystyle\int_{A}{\lambda(u)du}.
\end{equation}
We refer to $\lambda$ as \textit{intensity function}, or simply \textit{intensity}. In our framework, $N(A)$ is the random number of landslide detachment points inside $A$, where $A$ is an area of the valley. Then, $\nu$ is the expected number of landslide detachments occurring within $A$; we can reasonably assume that, as the measure of an area of the valley goes to 0, the risk of having a landslide detachment within this area is negligible. Hence, for the Radon-Nikodym theorem \cite{radnik}, the derivative of $\nu$, i.e. the intensity $\lambda$, exists. We are interested in modeling the intensity of the landslide detachment point process. In particular, we consider a \textit{spatial Poisson point process} on $W$ with intensity $\lambda$, i.e. a spatial process $X$ with domain, $W\subset \mathbb{R}^d$ such that:
\begin{enumerate}
   \item $N(A)\sim Poisson\biggl(\displaystyle\int_{A}{\lambda(u)d(u)}\biggr)$ for any $A$ compact subset of $W$.
   \item if $A_1$,...,$A_m$ are disjoint compact subsets of $W$, then $N(A_1)$,...,$N(A_m)$ are independent.
\end{enumerate}
~\\
When the intensity function $\lambda$ is constant in space, the process is called \textit{homogeneous}, and \textit{inhomogeneous} otherwise.
In the latter setting, let $z_i(u)$ denote the i-th covariate evaluated on $u \in W$, where $i=1,2\cdots q$.
We want to model $\lambda(u)$ as function of the geophysical covariates. The typical model for the intensity of an \textit{inhomogeneous Poisson process} is log-linear, and reads
\begin{equation}
log(\lambda(u))=\beta_0+\beta_1z_1(u)+\beta_2z_2(u)+\cdots+\beta_q z_q(u).
\label{linear_lambda}
\end{equation}
The model parameters are estimated via maximum likelihood \cite{coeurjoully19}. For an inhomogeneous Poisson point process with log-linear intensity, the log-likelihood function, up to a normalizing constant, is
\begin{equation}
    {l}(\beta)=\sum_{u\in X\cap W}{log(\lambda(u;\beta))}-\int_{W}{\lambda(u;\beta)du}
    \label{lik}
\end{equation}
where $\beta$ is the vector of the coefficients of the linear model, $W$ is the window, $X$ is the random set of points and $\lambda(u;\beta)$ is given by the right hand side of equation \eqref{linear_lambda}, as a function of $\beta$. For the log-linear model, one can use several penalty alternatives to standard maximum likelihood in order to regularize the problem and/or perform covariate selection \cite{yue_yu15}. When we consider a linear contribution of the covariates, the elastic net tool provides an infinite range of penalties from the pure regularization to the selection of covariates. In case of a spline regression, any kind of smoothing penalty can be added to reduce roughness or to select the effects. In settings where the true dependence between covariates and log-intensity is complex, it is not sufficient to include only the linear contributions of the covariates in the log-intensity model. In Appendix \ref{appendixA}, we report an argument motivating the convenience of including nonlinear contributions of the geophysical covariates. Employing a Generalized Additive Model (GAM \cite{hastie_tib_GAM}) to model the logarithmic intensity, the linear contribute associated to each covariate is substituted with a nonlinear contribute, for instance through cubic splines. Moreover, it yields benefits in enhancing the model flexibility while preserving interpretability. In GAM the penalty term used to perform regularization is the smoothing penalty \cite{wood17}. For our case study we smooth the continuous covariates with cubic regression splines and regularize with the smoothing splines penalty, i.e. the integral on $\mathbb{R}$ of the second derivative of the smoothing function. 

\subsection{Uncertainty quantification of the process intensity}
\label{semipar_boot}
To quantify the uncertainty related to the intensity estimation, we resort to a bootstrap procedure. In the work presented in \cite{bootPois}, the bootstrap replicates are set of points sampled from a spatial point process, whose intensity is estimated through a kernel estimator based on covariates. In the workflow proposed in this work, the GAM model stands as built-in method for covariate-based nonparametric estimation of the intensity. For this reason, as a straightforward alternative to the kernel estimated intensity of \cite{bootCov}, we propose the following semiparametric bootstrap procedure to quantify the uncertainty of the predicted intensity.

First, we use the fitted GAM model to estimate $\hat{\lambda}$, namely the landslide detachment map in the training valley. Then, for $B$ times we: (i) sample a new random set of points from the Poisson point process with intensity $\hat{\lambda}$, (ii) fit the model with this new set of points as training pattern, (iii) compute the intensity map from the fitted model. The resulting $B$ intensity maps make up the bootstrap sample and, for each pixel, the bootstrap standard deviation and the bootstrap percentile can be computed accordingly.


\section{Results and discussion}
\label{results}
In this section we report the results obtained by training the proposed models in Val Chiavenna, by selecting the model which best predicts the intensity in Upper Valtellina, and by testing it in Val Tartano to quantify its transferability. Notice that, in the context of our case study, the intensity function estimation provides the map of the density of landslide crowns per $m^2$. All analysis have been conducted by employing the R packages  \verb|spatstat| \cite{spatstat} and \verb|iRF| \cite{iRF}.

\subsection{Ranges of the covariates}
 \label{ranges}
\begin{figure}[H]
    \centering
    \includegraphics[scale=0.5]{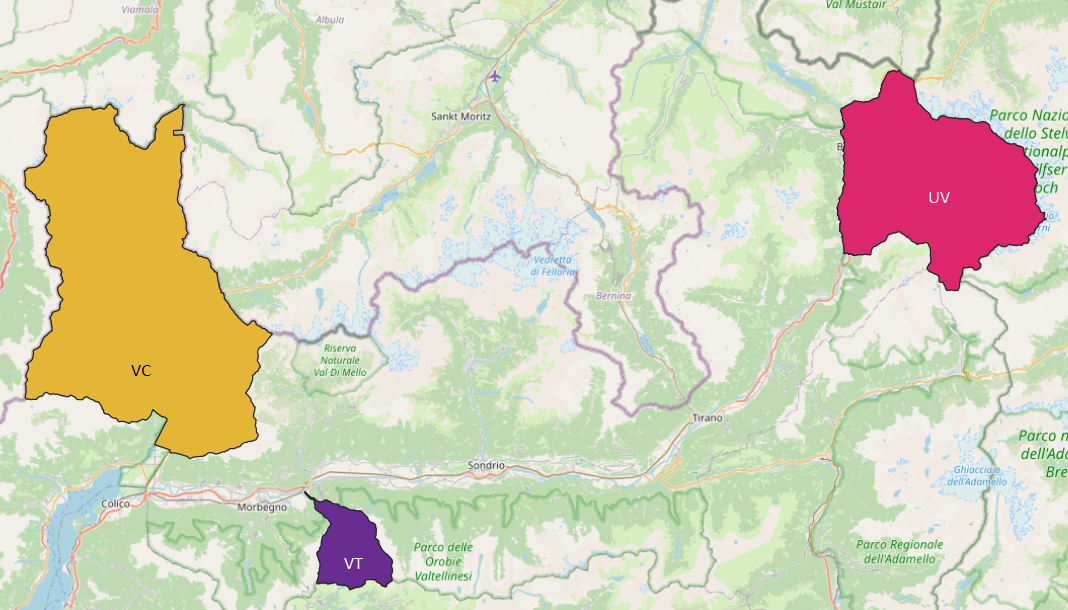}
    \caption{The areas covered by the three valleys in our dataset: from the left to the right Val Chiavenna, Val Tartano and Upper Valtellina. They are, respectively the training, testing and validation valleys.}
    \label{valleys}
\end{figure}

When considering the numerical geophysical covariates, it is important to analyze their variability within each valley. Since our goal is to develop a model that can be applied across different areas, selecting the appropriate training valley becomes crucial. During the training phase, the model should explore a diverse range of predictive covariates to minimize uncertainty when extrapolating to new valleys. Therefore, it is preferable for the training valley to exhibit the widest possible range of covariate values. As one can see from Figure \ref{valleys}, Val Chiavenna (VC) is the largest valley in terms of area.
\noindent
The geophysical characteristics within Val Tartano (VT) and Upper Valtellina (UV) exhibit marked differences in range. Specifically, UV boasts higher elevations, ranging from 1000 meters to 3820 meters, whereas VT maintains a lower elevation profile, everywhere being below 2500 meters. Additionally, UV experiences snow cover in select areas throughout the year, a phenomenon absent in VT, as indicated by the \textit{NDVI}. Conversely, VC displays large variability across key geophysical covariates such as \textit{DTM}, \textit{slope}, and \textit{NDVI}. As the ranges of these covariates include those observed within VT and UV, VC is regarded as the most suitable choice for training. Lastly, having UV a wider surface than VT, i.e. more pixels and then more data, UV is a suitable choice for the validation valley, since we favor an accurate choice of the model; hence VT is selected as test valley.

Let us focus on validation: we use maximum likelihood for model selection on UV. Comparing two models via log-likelihood (see Section \ref{method}) is valid only if computed on the same subarea. Also, for a reliable log-likelihood estimate, it must be calculated on an area where all covariates are in-range with respect to the training valley VC. Hence, during the selection phase, models are calibrated over the UV subarea where all covariates align with the training valley VC. To compare two models accurately, we select the intersection of subareas that are in-range for each model, ensuring log-likelihood estimations are both computed in the same area and in-range for both models.




\subsection{Covariates importance}
\label{cov_imp}
\begin{figure}[H]
    \centering
    \includegraphics[scale=0.4]{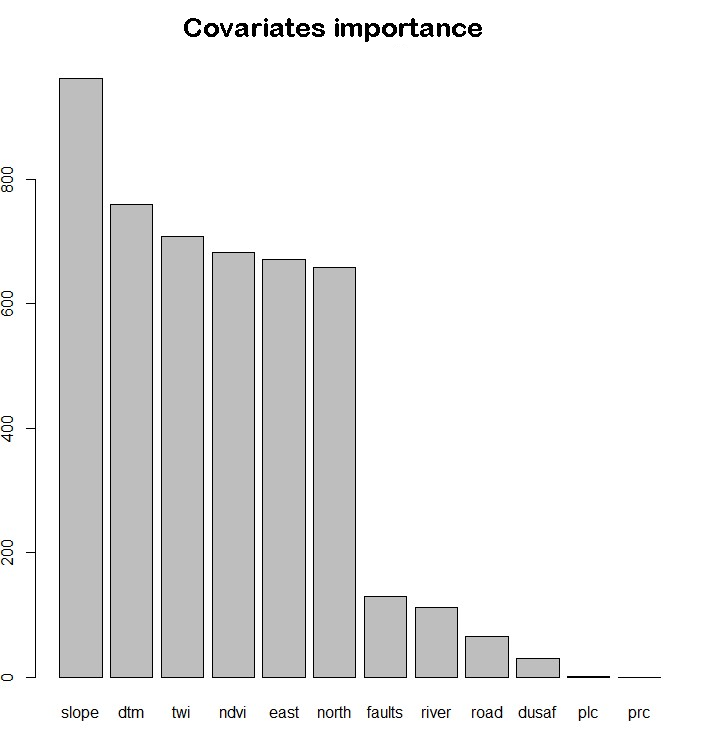}
    \caption{Covariates importance order in VC.}
    \label{irf_res}
\end{figure}

In order to drive the covariates selection for the GAM model of the logarithm of the intensity, we employ a Random Forest (RF) that is able to quantify the importance of the covariates. We train in VC a RF classifier, where the independent variables are the geophysical covariates and the dependent label is binary, taking value 1 if the pixel is a crown and 0 otherwise. Further details on the dataset used to train the RF are in  Appendix \ref{datairf}.  The metric used to order the covariates is the Gini importance. Figure \ref{irf_res} displays the importance of the geophysical covariates, retrieved using the RF.

The covariates are clearly divided into two blocks: the six most important covariates (i.e. \textit{slope}, \textit{DTM}, \textit{TWI}, \textit{NDVI}, \textit{east} and \textit{north}) and the least important ones (i.e. faults, rivers, roads, \textit{DUSAF}, \textit{PLC}, \textit{PRC}). \textit{Slope} is largely the most important covariate: this does not come as a surprise, since the object of our study are landslide crowns, that are the points from which the soil gives way under its own weight and that of the water accumulated from intense rains, and begins to slide down\cite{shallow_slope}. The covariate \textit{slope}, as expected, plays a huge role in the slip induction.

\subsection{Model selection}
\label{proposed_models}
Drawing on the covariates importance resulting from the RF, we adopt the following analytical scheme: (i) train the proposed models in VC, (ii) identify the best model, through a validation phase in UV, (iii) test the selected model in VT.
The first models suggested by the RF are the following generalized additive models for the logarithm of the intensity:
\begin{itemize}
    \item \textit{GAM-all}: keeping all the covariates as predictors.
    \item \textit{GAM-selected}: keeping only the 6 most important covariates.
\end{itemize}
These models arise from the considerations in Section \ref{cov_imp} and constitute a good baseline. However, an in-depth analysis of the impact of \textit{eastness} and \textit{northness}, extensively reported in Appendix \ref{appendix_modelliscartati}, suggests to consider a third model, in which \textit{northness} and \textit{eastness} are excluded from predictors. An analysis of the results shows that the latter model solves an underestimation problem that the first two had, confirming that the intuition to remove the two variables is correct. The third model is the following:
\begin{itemize}
    \item \textit{GAM-reduced}: keeping only the 4 most important covariates.
\end{itemize}
\noindent 
The GAM-reduced model results to be the one with highest likelihood in UV (see Table \ref{logliks}) among the proposed models, hence it is the selected model.

\begin{figure}[H]
    \centering
\includegraphics[scale=0.25]{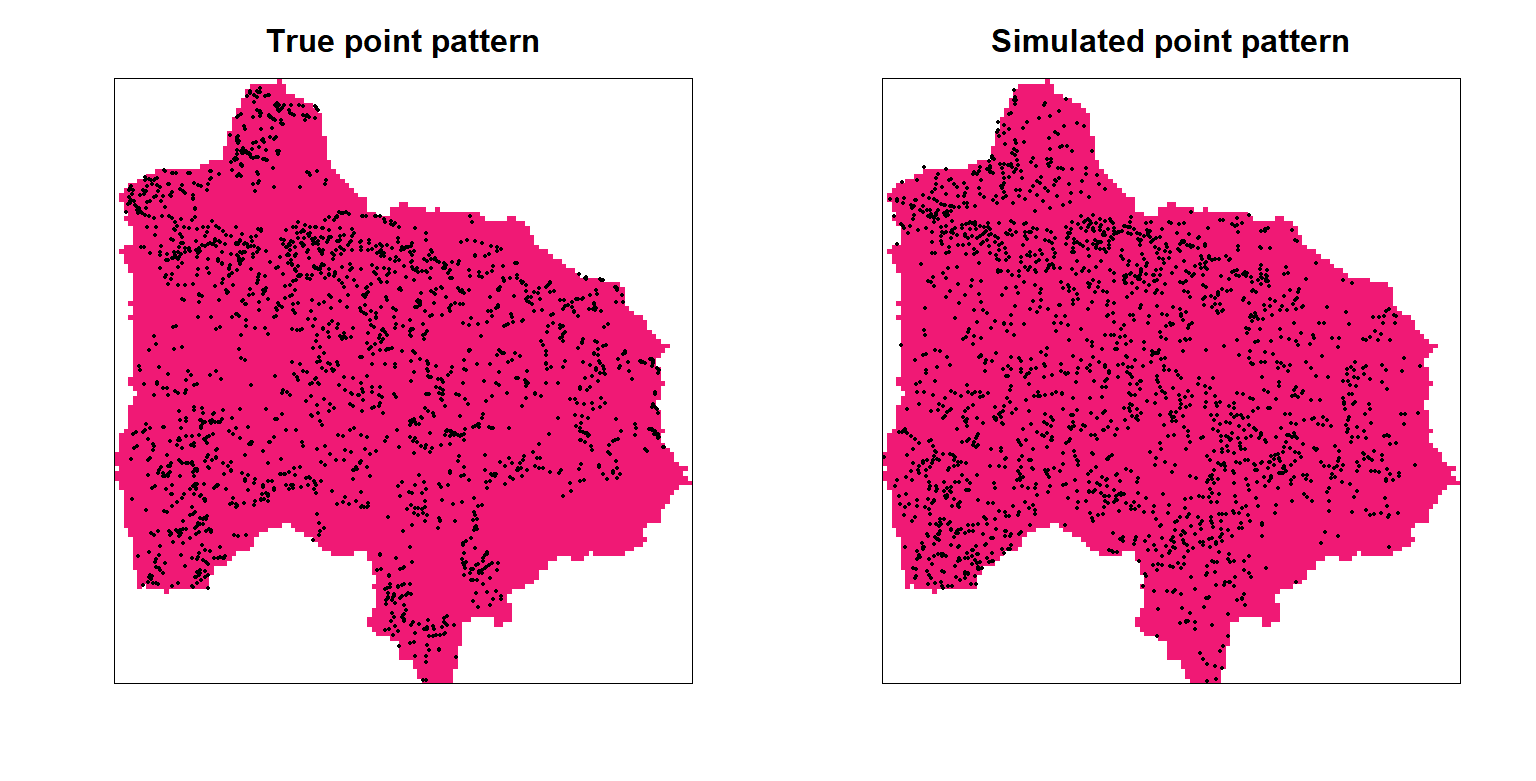}
    \caption{GAM-reduced with $TWI_b$: true pattern (left), a simulated pattern (right).}
    \label{Most4twib_sim}
\end{figure}

\begin{figure}[H]
    \centering
\includegraphics[scale=0.35]{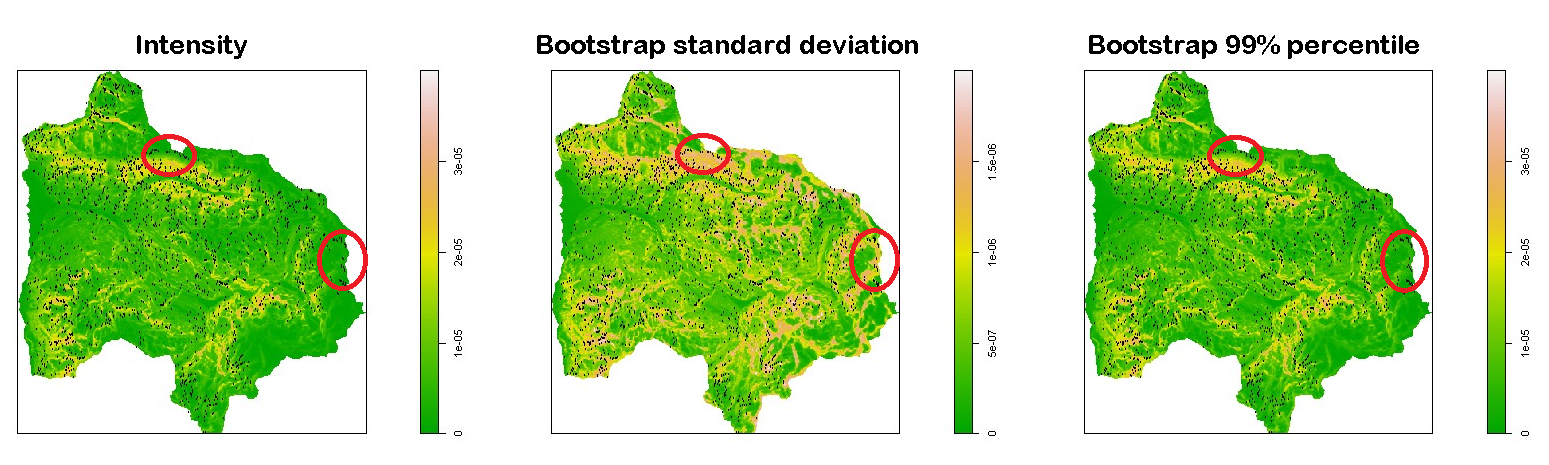}
    \caption{GAM-reduced with $TWI_b$: intensity (left), bootstrap sd (center), $99\%$-percentile (right). Enlighted two clouds of landslide crowns, unreported by the landslide detachment map, have a relatively high bootstrap standard deviation.}
    \label{Most4twib}
\end{figure}

\begin{table}[h]
\begin{tabular}{ll}
\toprule
Model & log-likelihood\\
\midrule
GAM-all & $-2.40\cdot 10^{4}$\\
GAM-selected & $-2.38\cdot 10^{4}$ \\
GAM-reduced & {$\boldsymbol{-2.34\cdot 10^{4}}$} \\
\botrule
\end{tabular}
\caption{Comparison of the proposed models through the log-likelihood}
\label{logliks}
\end{table}
In Figure \ref{Most4twib_sim} we report a simulation of point pattern sampled from the GAM-reduced with $TWI_b$, compared with the true one in UV. One can notice that the true (left) and the simulated (right) patterns present a strongly similar behaviour on the most of the surface of UV. The intensity can be considered as the primary indicator of activation risk. Nevertheless, in real applications the real intensity is unknown and only an estimate can be retrieved, with the reliability of the estimate possibly differing across locations. For this reason also the uncertainty of the intensity estimation has to be jointly taken into consideration. For example, if a zone of the valley is associated with a low intensity but the uncertainty of this estimation is high, the zone has to be considered potentially at risk. Hence the reason to report also the $99$th percentile map, i.e. the map that associates to each pixel the upper bound of the left-sided bootstrap confidence interval at level $99\%$. The $99$th percentile map plays the role of an \textit{alarm map}, since in addition to having high values where the estimated intensity is high, it has relatively high values even where the estimated intensity is low but there is high uncertainty of the estimate, so that it is unreliable. In Figure \ref{Most4twib}, we show the estimated intensity map in UV, the bootstrap standard deviation map and the $99$th percentile map. Some clouds of landslide crowns, unreported by the landslide detachment map, are the same that have a relatively high bootstrap standard deviation: the areas where they are located are likely characterized by combinations of geophysical properties that the model has not explored in the training set. The $99$th percentile map works well as alarm map as it reports these areas (see Figure \ref{Most4twib}).

\subsection{Model interpretation}
\begin{figure}[H]
    \centering
    \includegraphics[scale=0.3]{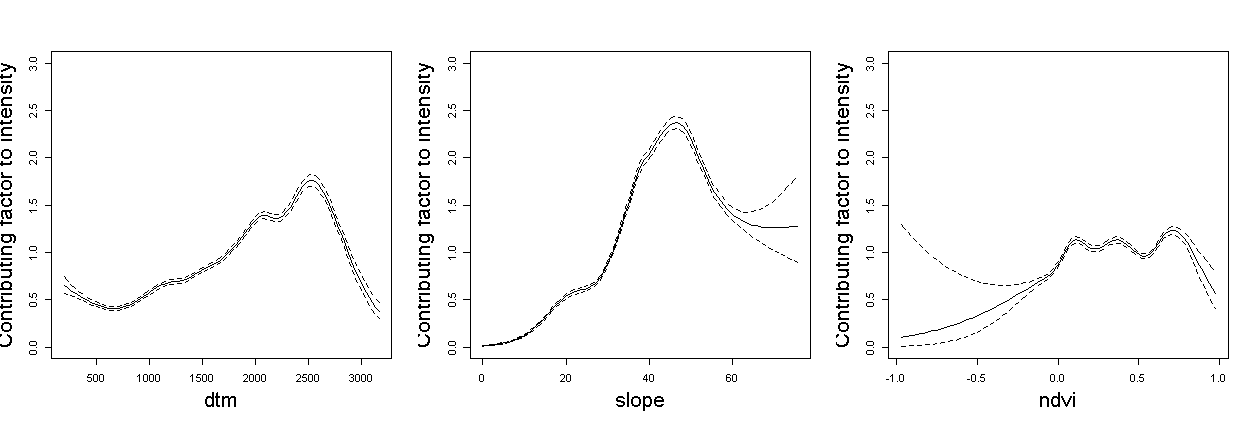}
    \caption{GAM-reduced model: contributing factors to intensity: DTM, slope and NDVI; the $TWI_b$ factor is equal to $exp(-0.18)\simeq0.835$.}
    \label{gam_effects}
\end{figure}

\begin{figure}[H]
    \centering   \includegraphics[scale=0.4]{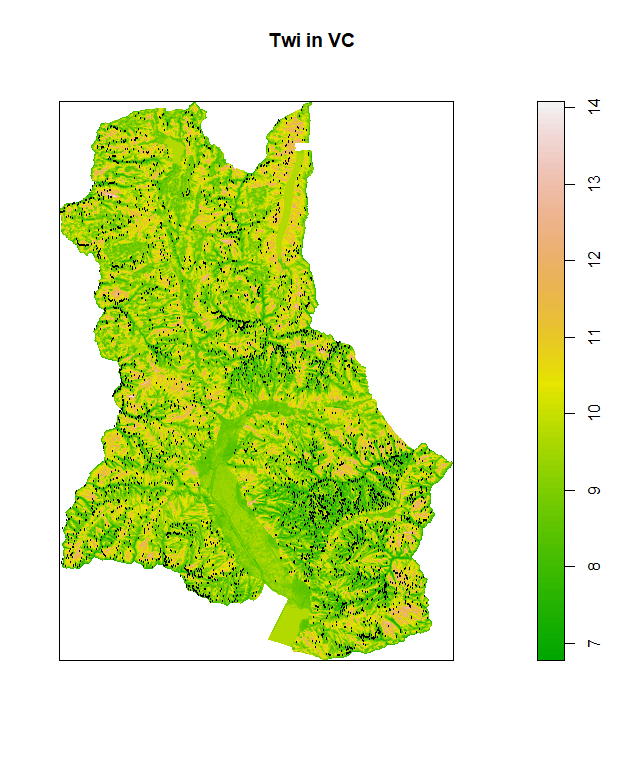}
    \caption{Topographic wetness index in Val Chiavenna. High values of TWI correspond to high potential ability to absorb water based on the topology (not on the soil type). Black dots mark the locations of the crowns of past landslides registered in the landslides inventory in VC.}
    \label{twi_VC}
\end{figure}

In the GAM-reduced model, the intensity is the product of the smoothed effects of $DTM$, $slope$ and $NDVI$, and the contribution factor of $TWI_b$. The model fitted in VC is characterized by the marginal effects displayed in Figure \ref{gam_effects}.
The high interpretability of the GAM model for the intensity offers now the opportunity to comment on the marginal effects of the covariates. First, we observe that the maximum effect of \textit{DTM} is obtained at around 2500 meters of elevation. The fact that, for higher values of altitude, higher values of elevation are associated with lower values of intensity can be due to multiple altitude-related factors, for instance soil type, temperature, and consequently precipitation amount and type, and so on. 
Now, let us focus on the effect of the \textit{slope}: for values less than 40°, the derivative is positive, until, between 40° and 60°, it reaches a maximum and it becomes negative. This behaviour is coherent with the nature of the landslides we study, i.e. \textit{shallow landslides}. Indeed, we a-priori know that, for high slopes, we expect falls, rather than shallow landslides.
In fact, for slopes greater than 45°, rocks are typically more stable due to the effect of gravity, so that collapses and breakages are more likely than shallow landslides \cite{slope_land}.
The estimated \textit{NDVI} effect tells that the higher the presence of water, the lower the intensity: areas subject to snow are more likely to be subject to avalanches, while the lake areas, as expected, are not subject to any kind of these events (neither landslides nor avalanches). Areas with high values of \textit{NDVI}, i.e. forests, are at lower risk, coherently with the well-known stabilizing effect of forests \cite{switz_shallow}.
Our model estimates the effect of \textit{TWI} as decreasing the risk of observing a landslide crown; this is also evident in Figure \ref{twi_VC}, where points are concentrated in areas with low values of \textit{TWI}. In the literature on the topic, high values of \textit{TWI} are known to increase landslide susceptibility \cite{suscept_korea} \cite{suscept_poland}. Nonetheless, we are not led to see this as a sign of model misspecification: indeed, it is crucial to stress that the probability of a landslide to originate at a given pixel deeply differs from its landslide susceptibility, i.e. the probability that the pixel is hit by a landslide. The fact that \textit{TWI} is a factor that decreases the intensity is not to be considered a contradiction. On the contrary, this result possibly means that a high \textit{TWI} is associated to areas that channel the flow of landslides, rather than to areas of activation of the sliding. In the next section we test the model in order to assess its transferability.

\subsection{Model testing}
\label{model_test}
\begin{figure}[H]
    \centering
\includegraphics[scale=0.3]{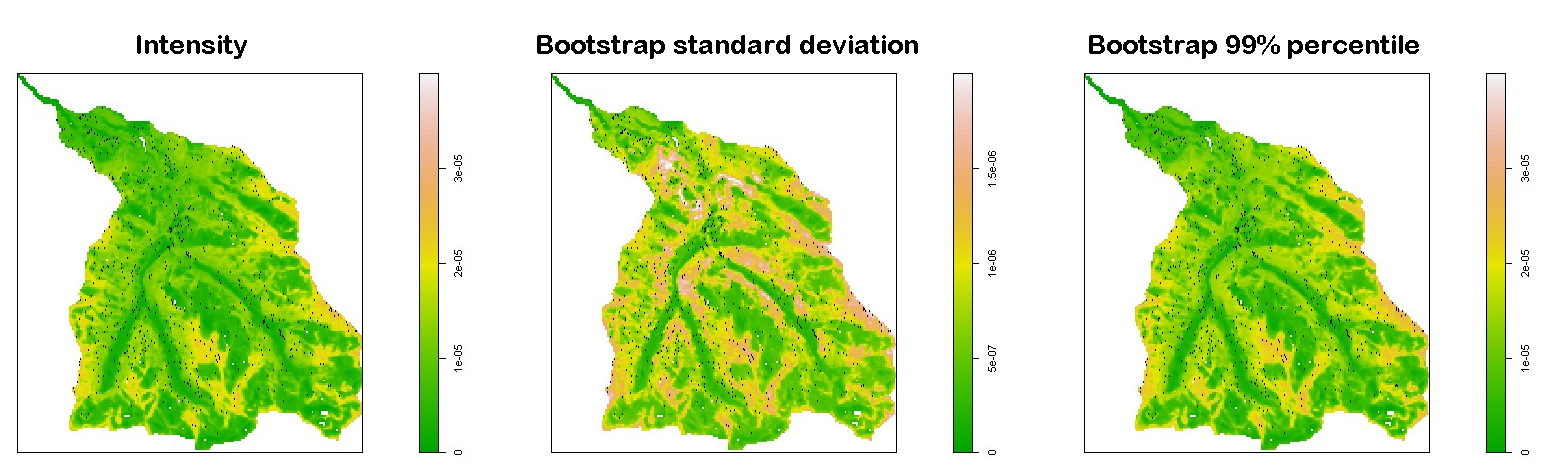}
    \caption{GAM-reduced with $TWI_b$: intensity (left), bootstrap sd (center), $99\%$-percentile (right). Some point clouds remain unmarked in the alarm map.}
    \label{Most4_VT}
\end{figure}

Likelihood plays the role of a comparison tool among models. Indeed, if we consider two models such that the first has a higher likelihood than the second, then the first model has a better fit to the observed data than the second one. However, the value of the likelihood alone does not assess the absolute goodness of the model in estimating the intensity on a new valley. In order to quantify and geographically localize possible mismatches between the fitted model and the observed data, we partitioned the testing valley into squared subareas and measured the so called \textit{raw error} of the model in each subarea, as defined in equation \ref{rawerreq}.
\begin{equation}
    e(A)=\widehat{\mathbb{E}[N(A)]}-n(A)
    \label{rawerreq}
\end{equation}
where $A$ is the subarea, $\widehat{\mathbb{E}[N(A)]}$ is the estimated expected number of landslides in $A$ according to the fitted model and $n$ the observed number of landslide crowns in $A$, according to the dataset. More explicitly:
\begin{equation}
    \mathbb{E}[N(A)]=\int_{A}{\hat{\lambda}(u)du}
\end{equation}
where $\hat{\lambda}$ is the intensity of the fitted model. The model can be judged on the basis of the summary statistics of the errors above (mean, median, variance, quartiles) and knowledge-based diagnostic can be performed in the subareas with largest errors, in order to understand information possibly missed by the model.
The predicted intensity, the standard deviation and the $99\%$-percentile maps, that result from employing the GAM-reduced model with $TWI_b$ in VT, are reported in Figure \ref{Most4_VT}. One can see that for most of the surface the model is able to detect the areas most at risk. In order to observe the raw errors, we consider a grid of $250m\times250m$ subareas, that partitions the test valley VT and allows one to compute regionalized residuals. This choice of the subareas' size allows one to better visualize the critical areas of our model in VT; however, this choice does not affect the interpretation of the residuals and might potentially derive from specific interests or prior expertise possessed by practitioners on the area under investigation.

\begin{figure}[H]
    \centering
\includegraphics[scale=0.3]{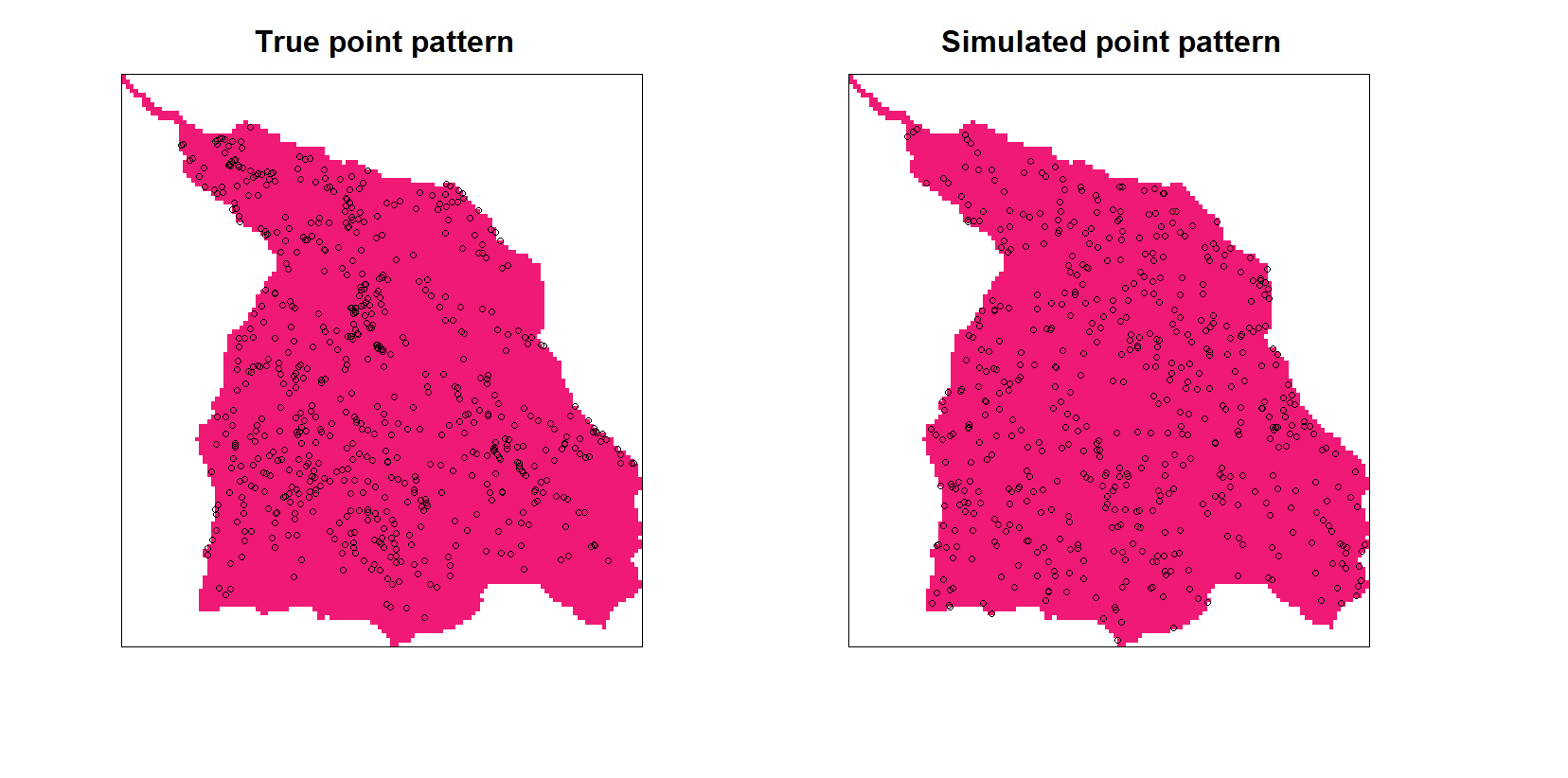}
    \caption{GAM-reduced with $TWI_b$: true pattern (left), a simulated pattern (right).}
    \label{Most4twib_sim_VT}
\end{figure} 

\begin{figure}[H]
    \centering
    \includegraphics[scale=0.35]{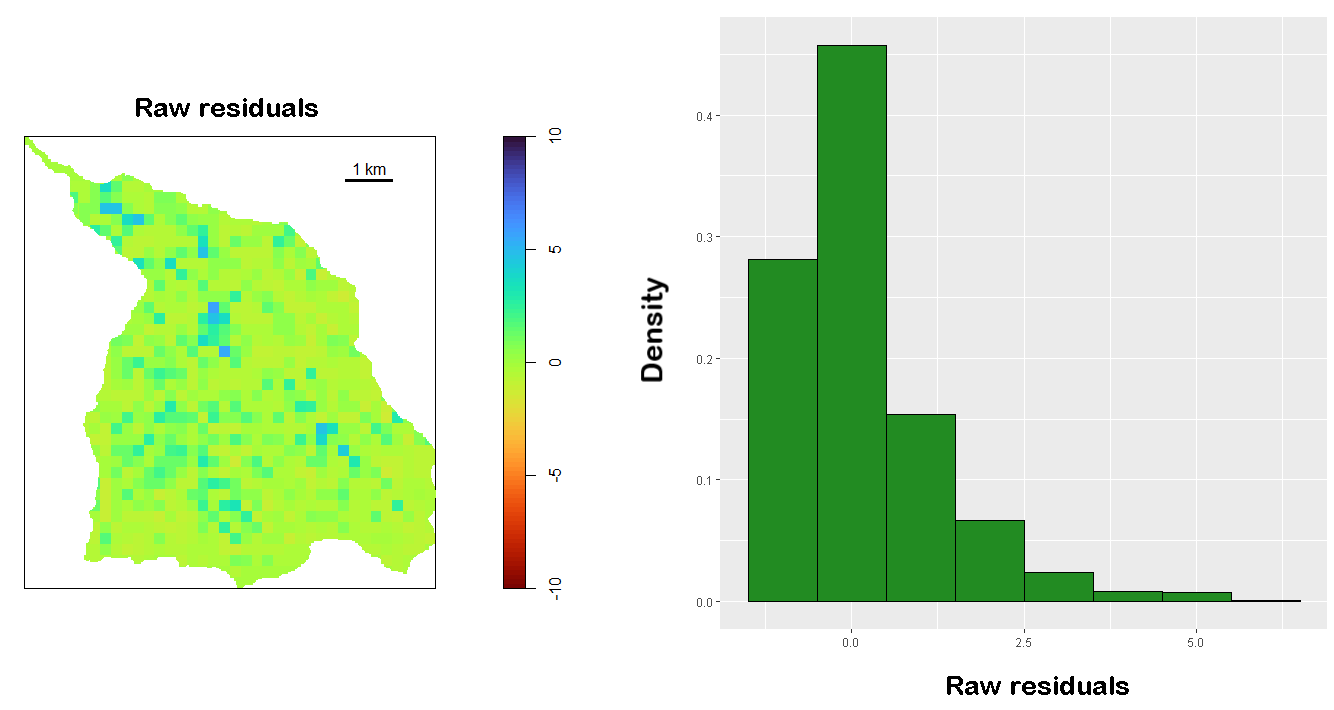}
    \caption{Raw errors,  i.e. difference between the observed number of crowns and its estimated expected number in each 250m x 250m area: one can notice some highly underestimated areas; however they are outliers with respect to the entire collection of areas.}
    \label{rawerrors}
\end{figure}

\begin{table}[h]
\begin{tabular}{@{}lllllll@{}}
\toprule
Vector & Min. & 1st Qu. & Median   & Mean & 3rd Qu. &  Max.\\
\midrule
Raw residuals & -1.2640 & -0.5374& -0.2197&  0.1432&  0.5246&  5.6370 \\
Abs. raw residuals & 0.0002 & 0.3140& 0.5360 &0.7321 & 0.7994 & 5.6370 \\
\botrule
\end{tabular}
\caption{Summary statistics on raw residuals}
\label{tableraw}
\end{table}

\noindent
Table \ref{tableraw} shows that, in general, the model works well on the test set: the third quartile of the raw residuals is considerably lower than the maximum value, implying that the reported underestimated subareas are potential outliers. The same holds for absolute value, as the $75\%$ of the time the absolute residual is lower than 1. In conclusion, the outcome of our testing is very positive in the perspective of transferability. 


\section{Conclusions}
\label{conclusion}
The prediction of shallow landslides is crucial for the preservation of lives and properties. A very informative point of this type of landslide is the crown, i.e. the pointwise approximation of the detachment area, in which intense and frequent rains trigger the sliding of debris and soil towards the valley. The identification of the relationship between the spatial distribution of landslide crowns and geophysical covariates makes it possible to understand which are the activating or stabilizing factors of landslides. The interpretability and flexibility of our models, obtained by employing generalized additive models (GAMs), allows us to observe in detail nonlinear relations between the intensity of the spatial point process and the covariates. In fact, with this approach, we have been able to widely discuss the obtained results. The Random Forest classifier plays a crucial role for the covariate selection, executed in order and block-wise. The estimate of the intensity of the landslide crowns process provides a map that represents the risk of landslide generation, then of alarm, and a tool to simulate landslide patterns. The bootstrap uncertainty quantification provides additional warning information: an area is at risk not only when the predicted intensity is high but also when the estimate, despite being low, has considerable uncertainty. The availability of data pertaining to three separate valleys enabled us to train, validate and test the model using disjoint datasets, thus making it possible to reliably assess its transferability. We found that our models, trained in VC, are able to recognise most of the risk zones of UV and VT and that they are able to quantify the number of landslides with high precision, based on available geophysical information. We also dealt with predictions where covariates are outside the range of the training dataset, since the estimate is expected to be possibly unreliable. To overcome this problem, we perform uncertainty quantification of the estimates. 
Possible under- or over-estimation are probably due to the fact that the weather conditions are not homogeneous in the area of investigation, hence in the future it would be interesting to include in the model the weather covariate. Having information on rainfalls and the date a landslide occurred would pave the way to understand how rain interacts with other covariates in triggering a landslide. For instance, we know that in 1987 a long series of landslides occurred, due to intense and frequent rainfalls in Valtellina that caused several landslide events within a few days \cite{article_sondrio}. Our model can be used for estimating and simulating shallow landslides in other valleys, in an enough similar context, for instance other valleys of the Alps of Lombardy or Switzerland. A model for the spatial distribution of the crowns can be combined with a model of the landslide dynamics in order to have a simulation of the landslide movements as a whole, from the origin to their dynamics towards the valley. Indeed, the landslides dynamics modelling is an hot topic in the today research (\cite{GATTI2024112798} \cite{GATTI2023105362} \cite{quecedo04} \cite{pastor17} \cite{pastor21} \cite{GATTI2024128525}). \\
Another possible further development can be to integrate the time with the space so that the covariates depend both on the position and on the date, thus modeling a space-time covariates-based process \cite{spatiotemp}. In this case, the temporal evolution of precipitation would be an information of fundamental importance, especially in the case of shallow landslides. Lastly, it can be interesting to apply our workflow in order to look for interpretable and transferable models of other classes of landslides or avalanches. \\

\bmhead{Acknowledgments}

This work is partially supported by ACCORDO Attuativo ASI-POLIMI “Attività di Ricerca e Innovazione” n. 2018-5-HH.0, collaboration agreement between the Italian Space Agency and Politecnico di Milano. \\
The work is partially funded by the Italian Ministry of Foreign Affairs and International Cooperation within the project “Geoinformatics and Earth Observation for Landslide Monitoring” - CUP D19C21000480001\\
GP, TB, and SV acknowledge the initiative “Dipartimento di Eccellenza 2023–2027”, MUR, Italy, Dipartimento di Matematica, Politecnico di Milano. \\
The authors gratefully acknowledge the financial support of IREA-CNR (Istituto per il Rilevamento Elettromagnetico dell'Ambiente del Consiglio Nazionale delle Ricerche) for funding a PhD grant in cooperation with Politecnico di Milano.

\begin{appendices}

\section{Justifying the use of GAM modelling the intensity}
\label{appendixA}
In this section we explain why we chose to model the intensity of the landslide crowns with a Generalized Additive Model (GAM) in which the candidate smoothed predictors are the continuous geophysical covariates \textit{DTM}, \textit{slope}, \textit{northness}, \textit{eastness}, \textit{TWI}, \textit{NDVI}. First we prove that the assumption of homogeneity is unrealistic, then we prove that the linear effect of the covariates is not sufficient to describe the dependence between them and the intensity.
\subsection{Homogeneity is unrealistic}
\begin{figure}[H]
    \centering
    \includegraphics[scale=0.25]{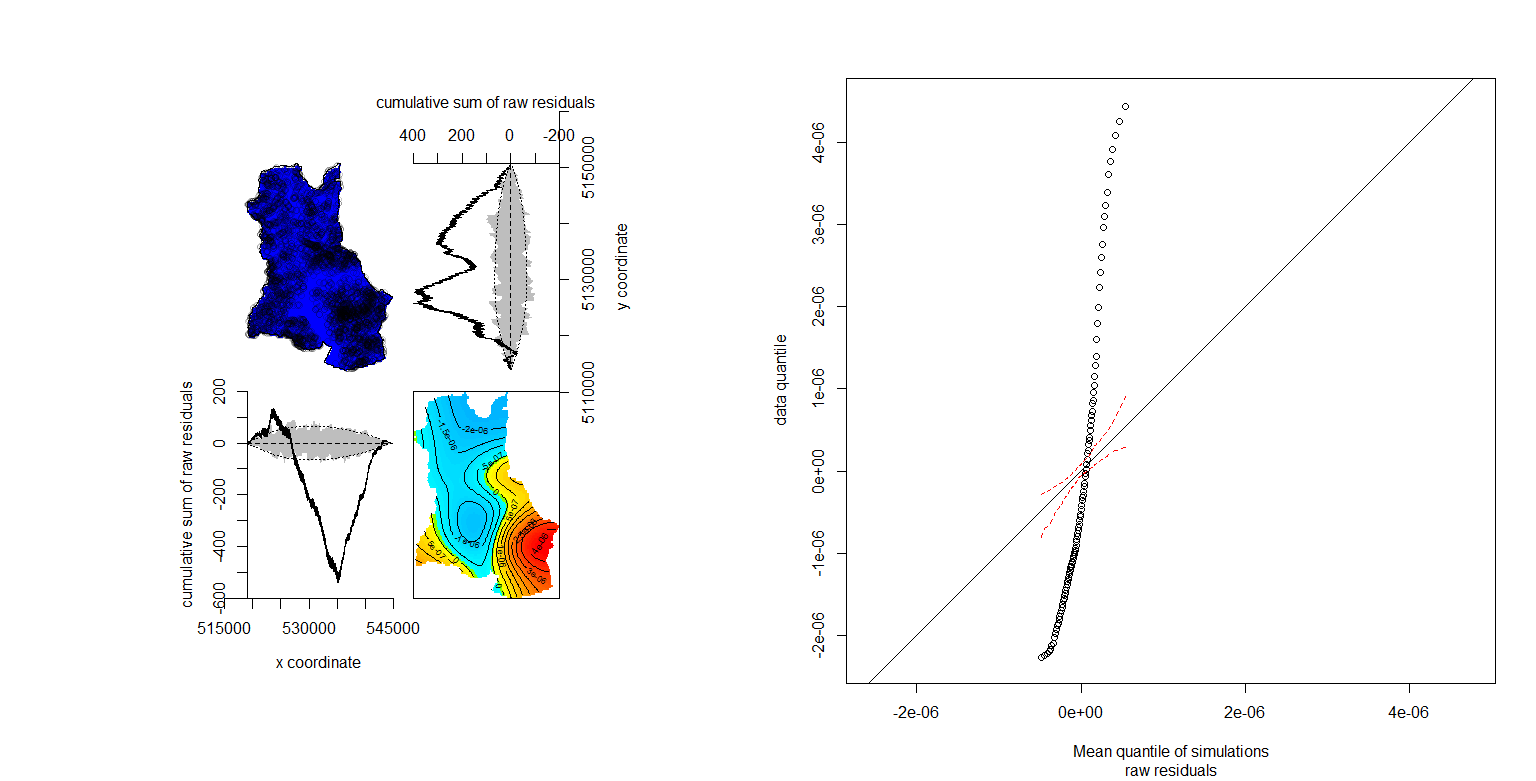}
    \caption{General diagnostic (left panel), QQ plot (right panel)}
    \label{diag_homo}
\end{figure}

Let's model the intensity assuming that it is spatially homogeneous. The value of the intensity function in the fitted model, trained in Val Chiavenna, is constant and equal to $7.45\cdot 10^{-6}$. Performing diagnostic \cite{baddeley05} \cite{baddeley07} on this model, we can see that the homogeneous model do not capture the spatial distribution (see Figure \ref{diag_homo}, left panel) of the intensity; moreover, if we do not condition the intensity to any covariate, the points are not independent of each other and then the points are unlikely distributed as a Spatial Poisson Point Process (see Figure \ref{diag_homo}, right panel).

\begin{figure}[H]
    \centering
    \includegraphics[scale=0.25]{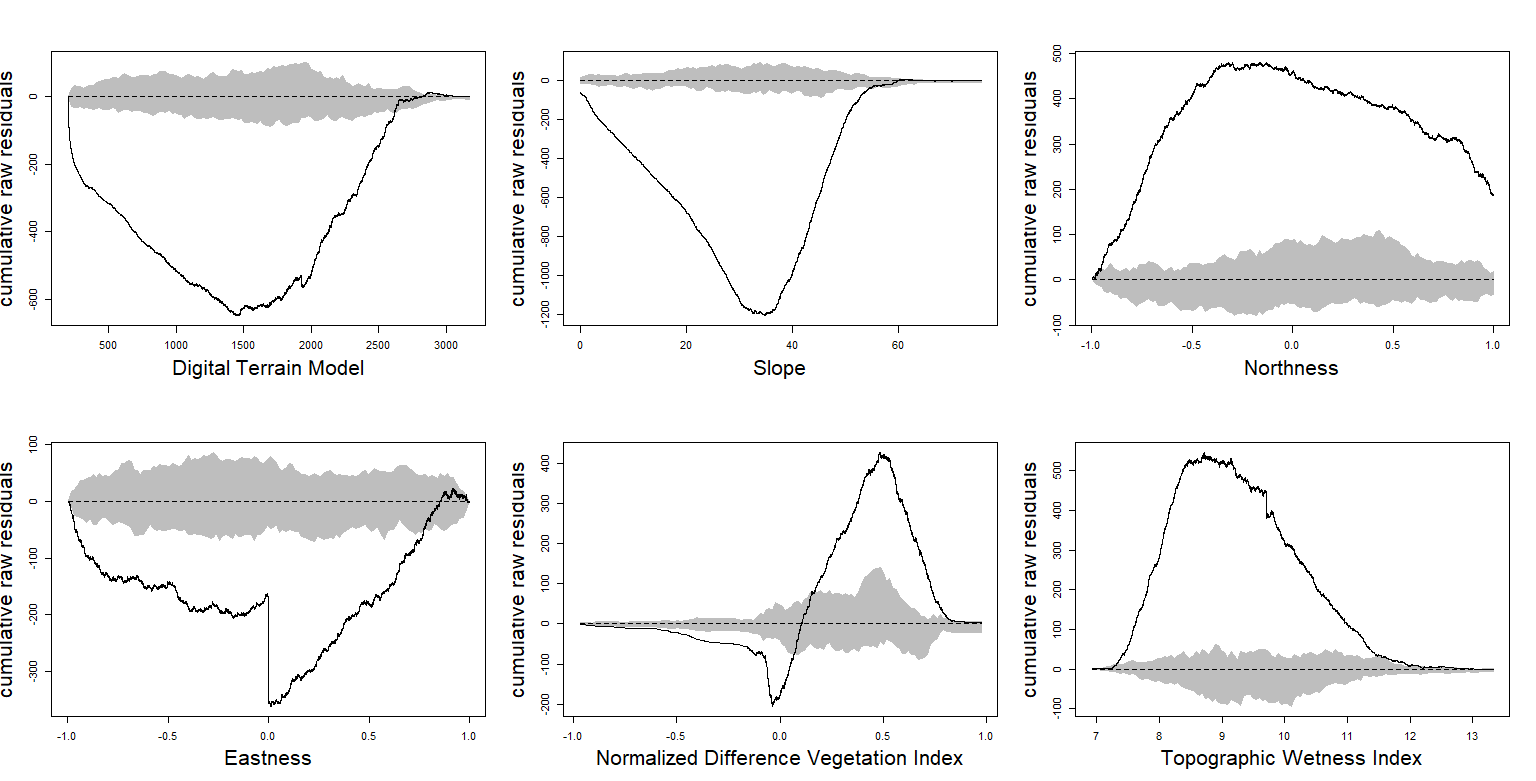}
    \caption{Lurking covariate plots}
    \label{lurking_homo}
\end{figure}

In particular, checking the lurking covariate plots on the continuous covariates, we can clearly notice that they are necessary to model the intensity: indeed, the cumulative raw residual curve is out of the bootstrap confidence bands (see Figure \ref{lurking_homo}).

\subsection{Nonlinear effects of the covariates are needed}
Once we know that the geophysical continuous covariates are necessary in modeling the intensity of the landslide crowns process, we wonder if it is sufficient to consider their linear contribution. Now, we fit, in Val Chiavenna, the following linear model:

\begin{equation}  \footnotesize\lambda(u)=\beta_0+\beta_{d}DTM(u)+\beta_{s}slope(u)+\beta_{E}east(u)+\beta_{N}north(u)+\beta_{t}TWI(u)+\beta_{n}NDVI(u)
\label{eq_linear}
\end{equation}

\begin{figure}[H]
    \centering
    \includegraphics[scale=0.25]{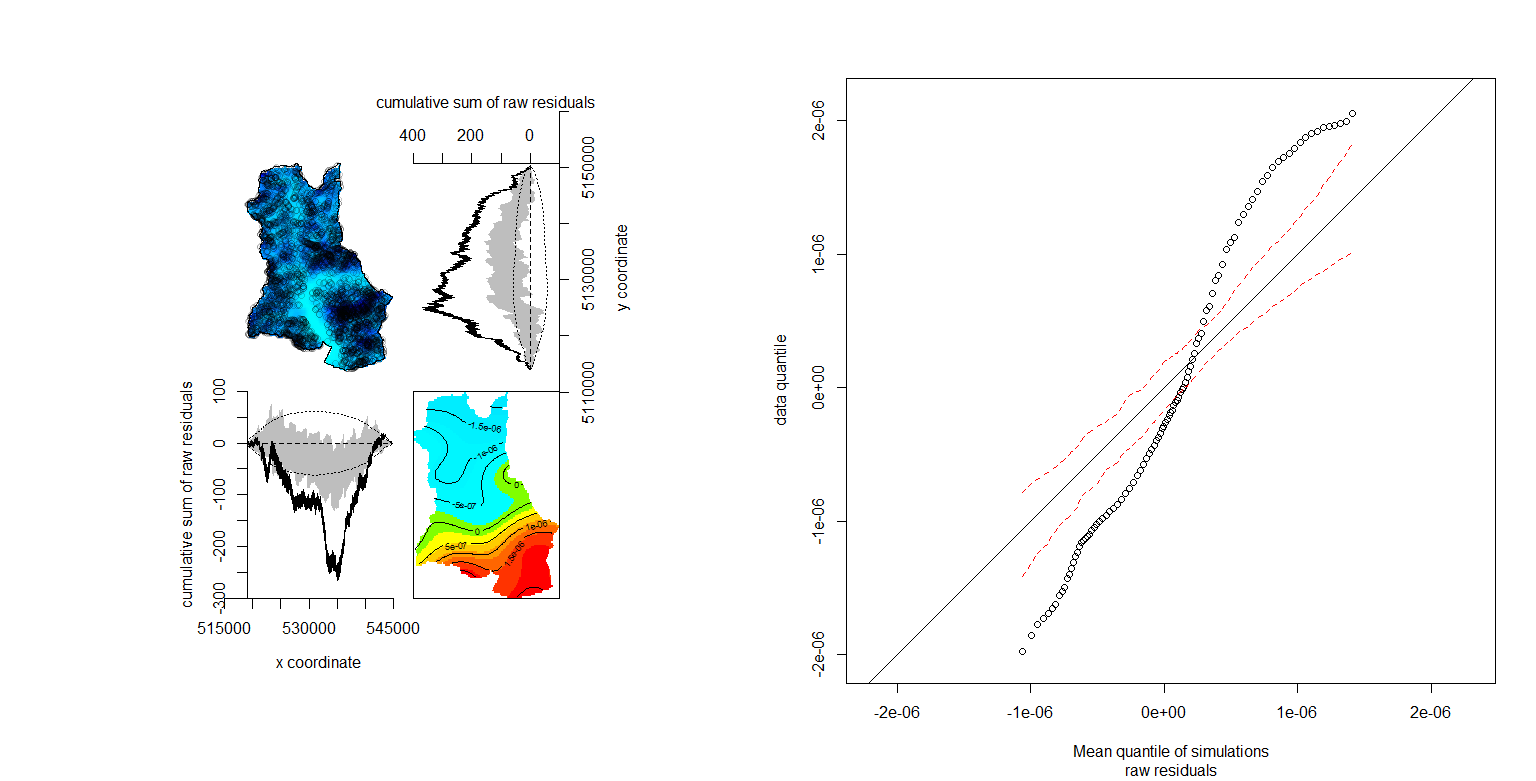}
    \caption{General diagnostic (left panel), QQ plot (right panel)}
    \label{diag_lincov}
\end{figure}

\begin{figure}[H]
    \centering
    \includegraphics[scale=0.25]{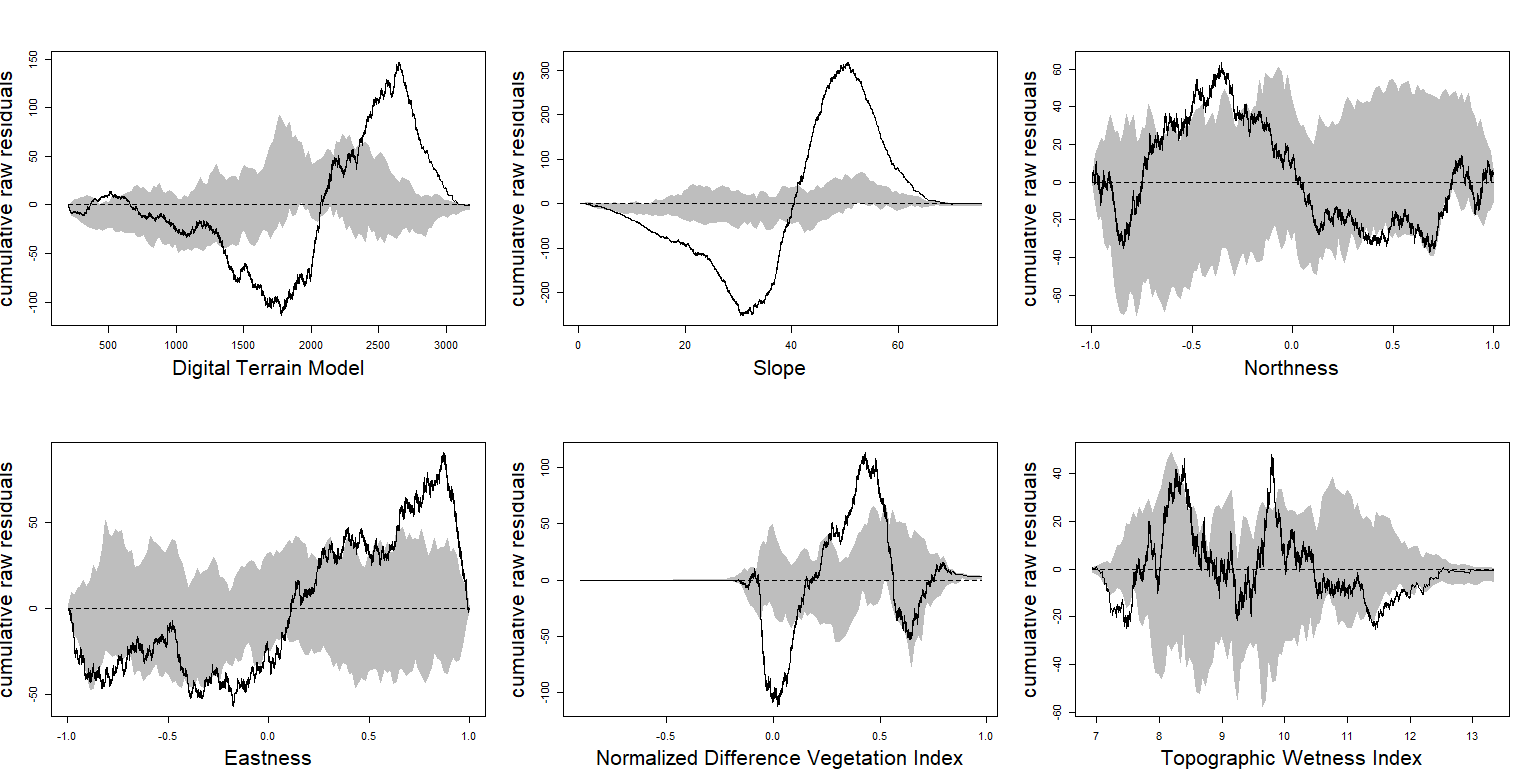}
    \caption{Lurking covariate plots}
    \label{lurking_lincov}
\end{figure}

\noindent The diagnostic reveals an improvement in the model goodness-of-fit by considering the covariates; however, the model has still some criticalities: spatial dependence is not completely captured (see Figure \ref{diag_lincov}, left panel), the model, as written in the Equation \ref{eq_linear}, would be improved by considering the interaction between points (see Figure \ref{diag_lincov}, right panel), and it is evident that a linear effect is not sufficient to capture the true relation between intensity and covariates, especially for \textit{DTM}, \textit{slope} and \textit{NDVI} (see Figure \ref{lurking_lincov}. This result prove that we need to consider a nonlinear contribution, hence a flexible and explainable model for the intensity: GAM.

\section{Dataset used to train RF}
\label{datairf}
\begin{figure}[H]
    \centering
    \includegraphics[scale=0.35]{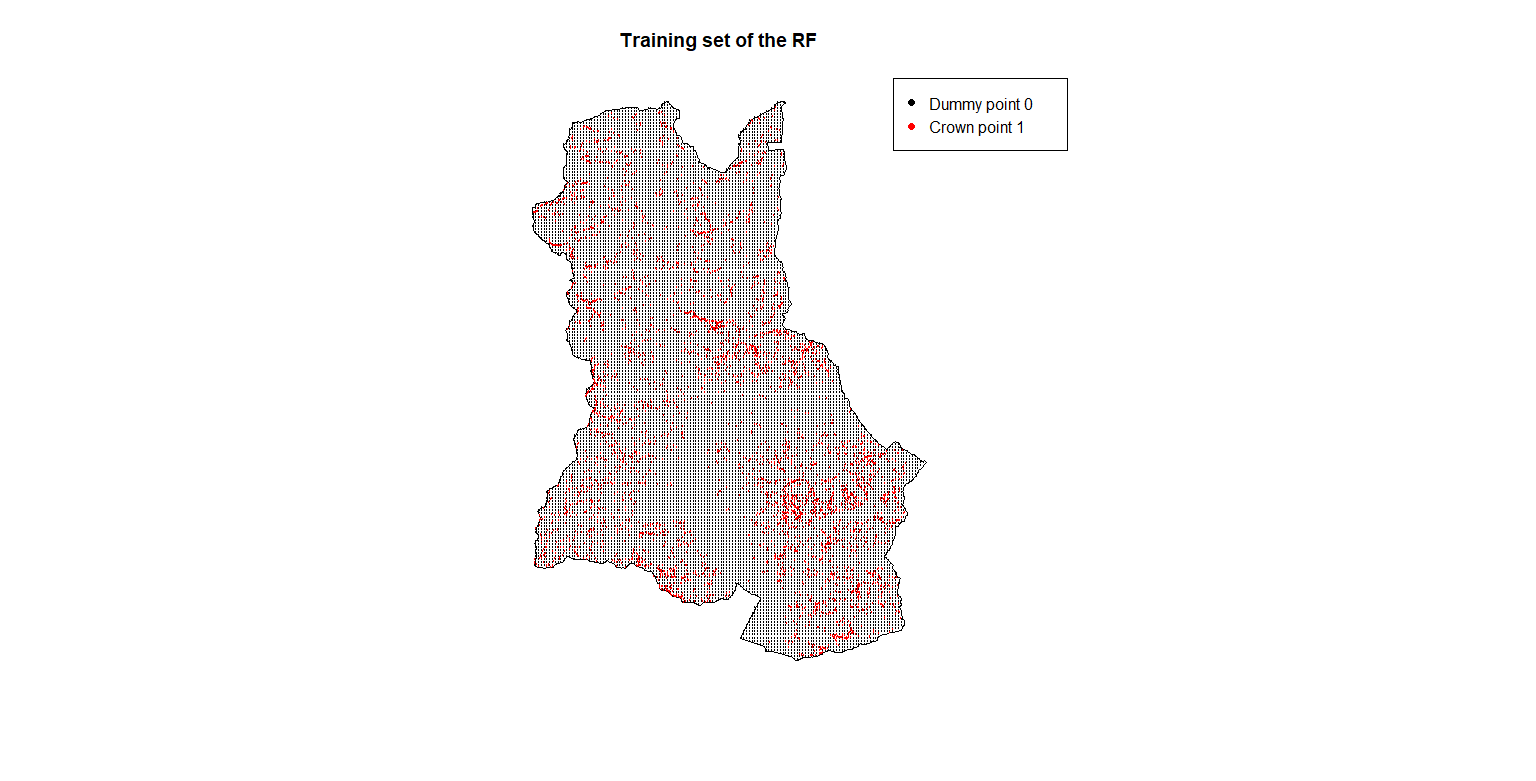}
    \caption{RF dataset (Val Chiavenna)}
    \label{irfdata2}
\end{figure}

In Figure \ref{irfdata2} we show the training outcomes (on VC) of the trained RF; the black points, called \textit{dummy points} and positioned in a regular and dense grid into the VC surface, constitute the class 0 (no landslide crown), while the red points are the positions of the landslide crowns recorded in VC and they constitute the class 1. The trained RF is a binary classifier whose input are the geophysical covariates and the output is the binary class, 0 if no landslide crown, 1 if landslide crown.

\section{GAM-selected: underestimation of the intensity due to orientation}
\label{appendix_modelliscartati}
\begin{figure}[H]
    \centering
    \includegraphics[scale=0.25]{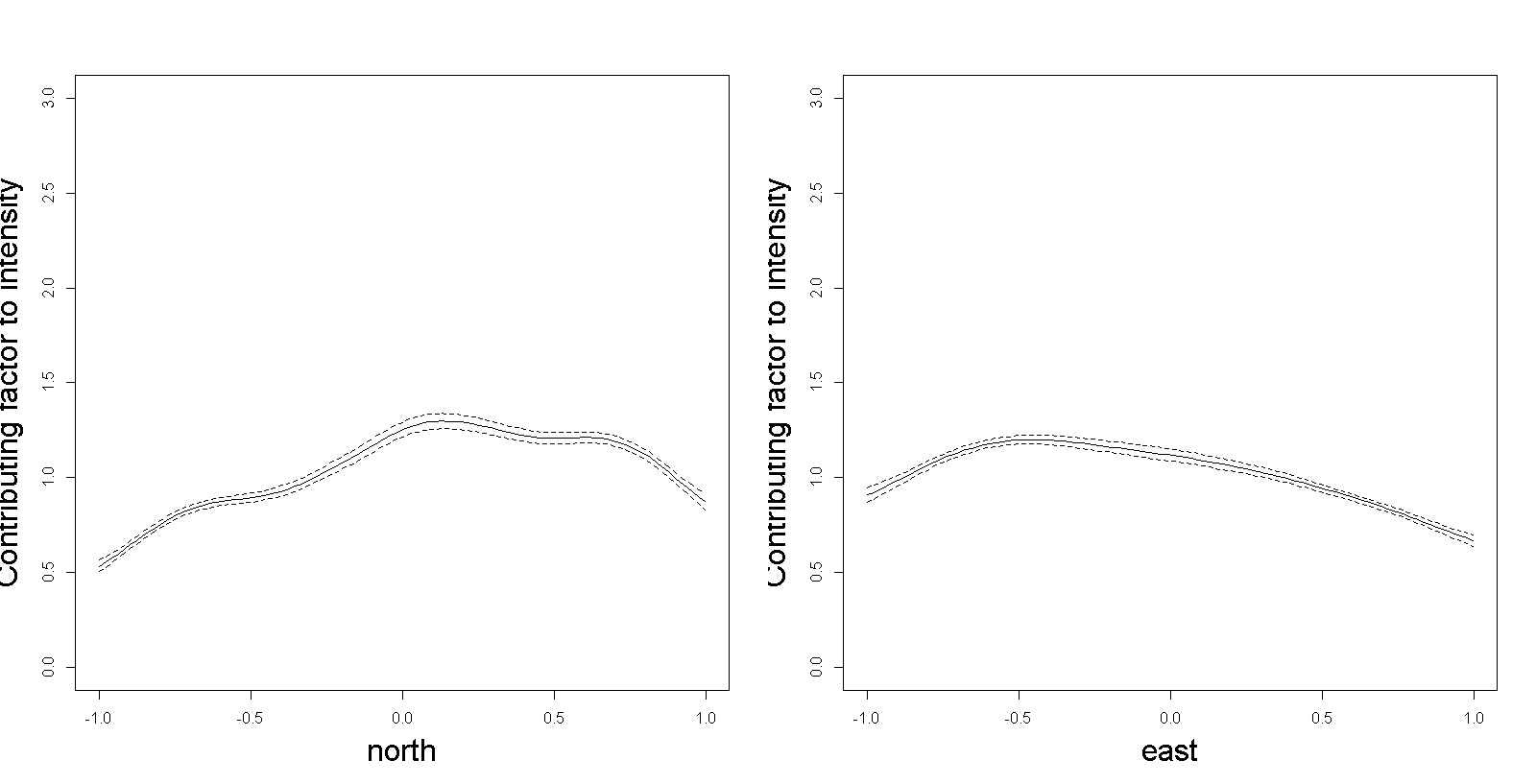}
    \caption{Contributing factors of \textit{eastness} and \textit{northness} to intensity}
    \label{eff_orient}
\end{figure}

\begin{figure}[H]
    \centering
    \includegraphics[scale=0.6]{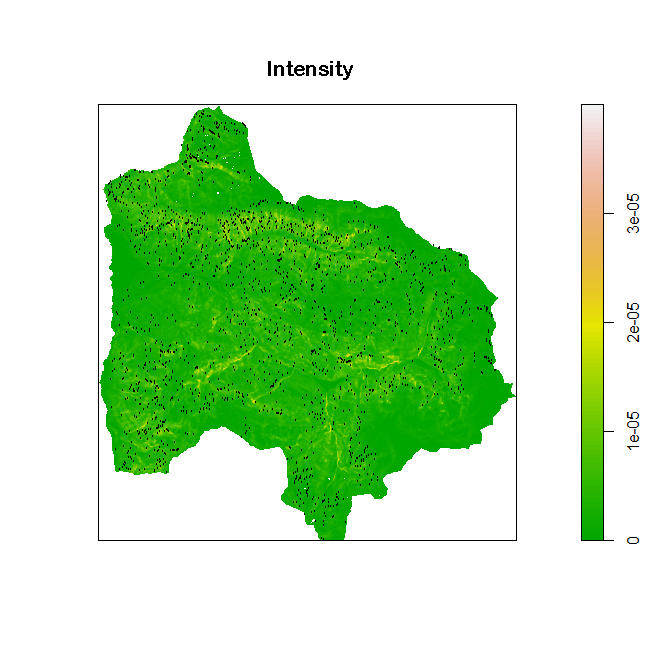}
    \caption{GAM-selected: intensity in UV}
    \label{pred_6}
\end{figure}

\begin{figure}[H]
    \centering
    \includegraphics[scale=0.4]{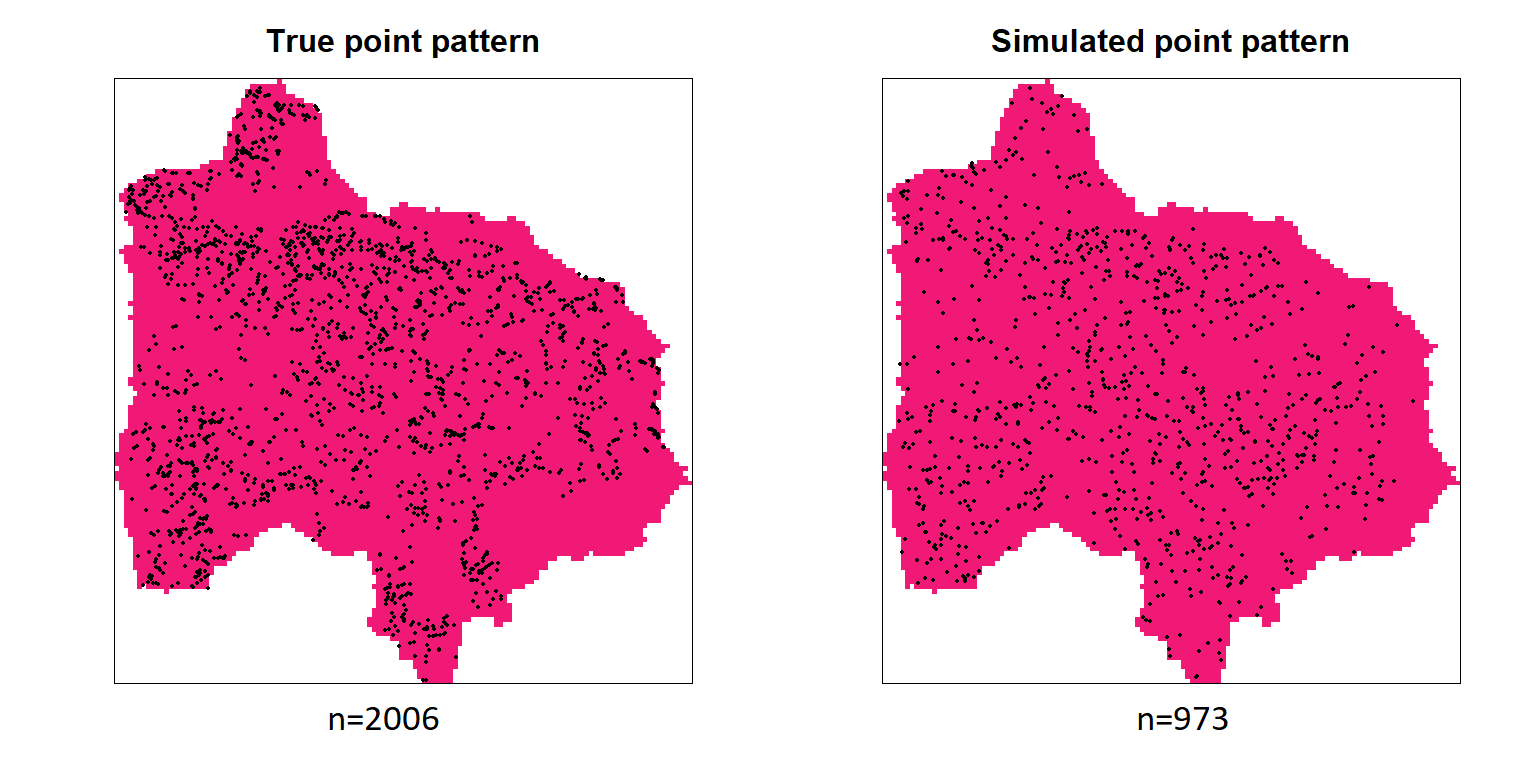}
    \caption{GAM-selected: true pattern (left), simulated pattern (right)}
    \label{sim_6}
\end{figure}

In this section we report the results obtained by fitting the GAM-selected model in VC (the results for GAM-All are strongly similar). In Figure \ref{eff_orient} one can observe that the GAM effects of \textit{eastness} and \textit{northness} do not show any clear and remarkable behaviour. By employing the model in UV, it is evident, from Figure \ref{pred_6} and \ref{sim_6}, that the intensity is highly underestimated; indeed, it predicts less landslides than half of the true number of landslides in UV. At this stage of the work, the RF was crucial, as it suggests that \textit{northness} and \textit{eastness} are the least important among the predictors of the GAM-selected model, so by eliminating them from the model we quickly solved the problem of underestimating intensity.

\end{appendices}


\bibliography{sn-bibliography}


\begin{thebibliography}{40}
\ifx \bisbn   \undefined \def \bisbn  #1{ISBN #1}\fi
\ifx \binits  \undefined \def \binits#1{#1}\fi
\ifx \bauthor  \undefined \def \bauthor#1{#1}\fi
\ifx \batitle  \undefined \def \batitle#1{#1}\fi
\ifx \bjtitle  \undefined \def \bjtitle#1{#1}\fi
\ifx \bvolume  \undefined \def \bvolume#1{\textbf{#1}}\fi
\ifx \byear  \undefined \def \byear#1{#1}\fi
\ifx \bissue  \undefined \def \bissue#1{#1}\fi
\ifx \bfpage  \undefined \def \bfpage#1{#1}\fi
\ifx \blpage  \undefined \def \blpage #1{#1}\fi
\ifx \burl  \undefined \def \burl#1{\textsf{#1}}\fi
\ifx \doiurl  \undefined \def \doiurl#1{\url{https://doi.org/#1}}\fi
\ifx \betal  \undefined \def \betal{\textit{et al.}}\fi
\ifx \binstitute  \undefined \def \binstitute#1{#1}\fi
\ifx \binstitutionaled  \undefined \def \binstitutionaled#1{#1}\fi
\ifx \bctitle  \undefined \def \bctitle#1{#1}\fi
\ifx \beditor  \undefined \def \beditor#1{#1}\fi
\ifx \bpublisher  \undefined \def \bpublisher#1{#1}\fi
\ifx \bbtitle  \undefined \def \bbtitle#1{#1}\fi
\ifx \bedition  \undefined \def \bedition#1{#1}\fi
\ifx \bseriesno  \undefined \def \bseriesno#1{#1}\fi
\ifx \blocation  \undefined \def \blocation#1{#1}\fi
\ifx \bsertitle  \undefined \def \bsertitle#1{#1}\fi
\ifx \bsnm \undefined \def \bsnm#1{#1}\fi
\ifx \bsuffix \undefined \def \bsuffix#1{#1}\fi
\ifx \bparticle \undefined \def \bparticle#1{#1}\fi
\ifx \barticle \undefined \def \barticle#1{#1}\fi
\bibcommenthead
\ifx \bconfdate \undefined \def \bconfdate #1{#1}\fi
\ifx \botherref \undefined \def \botherref #1{#1}\fi
\ifx \url \undefined \def \url#1{\textsf{#1}}\fi
\ifx \bchapter \undefined \def \bchapter#1{#1}\fi
\ifx \bbook \undefined \def \bbook#1{#1}\fi
\ifx \bcomment \undefined \def \bcomment#1{#1}\fi
\ifx \oauthor \undefined \def \oauthor#1{#1}\fi
\ifx \citeauthoryear \undefined \def \citeauthoryear#1{#1}\fi
\ifx \endbibitem  \undefined \def \endbibitem {}\fi
\ifx \bconflocation  \undefined \def \bconflocation#1{#1}\fi
\ifx \arxivurl  \undefined \def \arxivurl#1{\textsf{#1}}\fi
\csname PreBibitemsHook\endcsname

\bibitem[\protect\citeauthoryear{Loche et~al.}{2022}]{loche2022}
\begin{barticle}
\bauthor{\bsnm{Loche}, \binits{M.}},
\bauthor{\bsnm{Alvioli}, \binits{M.}},
\bauthor{\bsnm{Marchesini}, \binits{I.}},
\bauthor{\bsnm{Haakan}, \binits{B.}},
\bauthor{\bsnm{Lombardo}, \binits{L.}}:
\batitle{Landslide susceptibility maps of italy: Lesson learnt from dealing
  with multiple landslide types and the uneven spatial distribution of the
  national inventory}.
\bjtitle{Earth-Science Reviews}
\bvolume{232 (2022)},
\bfpage{1}--\blpage{21}
(\byear{2022})
\doiurl{10.1016/j.earscirev.2022.104125}
\end{barticle}
\endbibitem

\bibitem[\protect\citeauthoryear{Conforti and Ietto}{2021}]{land_sus_validated}
\begin{barticle}
\bauthor{\bsnm{Conforti}, \binits{M.}},
\bauthor{\bsnm{Ietto}, \binits{F.}}:
\batitle{Modeling shallow landslide susceptibility and assessment of the
  relative importance of predisposing factors, through a gis-based statistical
  analysis}.
\bjtitle{Geosciences}
\bvolume{11},
\bfpage{333}
(\byear{2021})
\doiurl{10.3390/geosciences11080333}
\end{barticle}
\endbibitem

\bibitem[\protect\citeauthoryear{Oguz et~al.}{2021}]{article_SL1}
\begin{botherref}
\oauthor{\bsnm{Oguz}, \binits{E.A.}},
\oauthor{\bsnm{Depina}, \binits{I.}},
\oauthor{\bsnm{Thakur}, \binits{V.}}:
Effects of soil heterogeneity on susceptibility of shallow landslides.
Landslides
\textbf{19}
(2021)
\doiurl{10.1007/s10346-021-01738-x}
\end{botherref}
\endbibitem

\bibitem[\protect\citeauthoryear{Wang et~al.}{2023}]{article_SL2}
\begin{barticle}
\bauthor{\bsnm{Wang}, \binits{J.}},
\bauthor{\bsnm{Gong}, \binits{Q.}},
\bauthor{\bsnm{Yuan}, \binits{S.}},
\bauthor{\bsnm{Chen}, \binits{J.}}:
\batitle{Combining soil macropore flow with formation mechanism to the
  development of shallow landslide warning threshold in south china}.
\bjtitle{Frontiers in Earth Science}
\bvolume{10},
\bfpage{1048427}
(\byear{2023})
\doiurl{10.3389/feart.2022.1048427}
\end{barticle}
\endbibitem

\bibitem[\protect\citeauthoryear{Ballabio and Sterlacchini}{2012}]{SVM_SL}
\begin{botherref}
\oauthor{\bsnm{Ballabio}, \binits{C.}},
\oauthor{\bsnm{Sterlacchini}, \binits{S.}}:
Support vector machines for landslide susceptibility mapping: The staffora
  river basin case study, italy.
Mathematical geosciences
\textbf{44}
(2012)
\doiurl{10.1007/s11004-011-9379-9}
\end{botherref}
\endbibitem

\bibitem[\protect\citeauthoryear{Ha et~al.}{2020}]{RF_SL}
\begin{barticle}
\bauthor{\bsnm{Ha}, \binits{N.V.}},
\bauthor{\bsnm{Ataollah}, \binits{S.}},
\bauthor{\bsnm{Himan}, \binits{S.}},
\bauthor{\bsnm{Wei}, \binits{C.}},
\bauthor{\bsnm{John}, \binits{C.}},
\bauthor{\bsnm{Marten}, \binits{G.}},
\bauthor{\bsnm{Abolfazl}, \binits{J.}},
\bauthor{\bsnm{Mohammadtaghi}, \binits{A.}},
\bauthor{\bsnm{Shaghayegh}, \binits{M.}},
\bauthor{\bsnm{Davood}, \binits{A.}},
\bauthor{\bsnm{Binh}, \binits{P.}},
\bauthor{\bsnm{Bin}, \binits{A.B.}},
\bauthor{\bsnm{Ahmad}},
\bauthor{\bsnm{Saro}, \binits{L.}}:
\batitle{Shallow landslide susceptibility mapping by random forest base
  classifier and its ensembles in a semi-arid region of iran}.
\bjtitle{Forests}
\bvolume{11},
\bfpage{421}
(\byear{2020})
\doiurl{10.3390/f11040421}
\end{barticle}
\endbibitem

\bibitem[\protect\citeauthoryear{Wu et~al.}{2020}]{Adaboost_SL_validno}
\begin{barticle}
\bauthor{\bsnm{Wu}, \binits{Y.}},
\bauthor{\bsnm{Ke}, \binits{Y.}},
\bauthor{\bsnm{Chen}, \binits{Z.}},
\bauthor{\bsnm{Liang}, \binits{S.}},
\bauthor{\bsnm{Zhao}, \binits{H.}},
\bauthor{\bsnm{Hong}, \binits{H.}}:
\batitle{Application of alternating decision tree with adaboost and bagging
  ensembles for landslide susceptibility mapping}.
\bjtitle{Catena}
\bvolume{187},
\bfpage{104396}
(\byear{2020})
\doiurl{10.1016/j.catena.2019.104396}
\end{barticle}
\endbibitem

\bibitem[\protect\citeauthoryear{Xiong et~al.}{2022}]{CNN_SL}
\begin{barticle}
\bauthor{\bsnm{Xiong}, \binits{Y.}},
\bauthor{\bsnm{Zhou}, \binits{Y.}},
\bauthor{\bsnm{Wang}, \binits{F.}},
\bauthor{\bsnm{Wang}, \binits{S.}},
\bauthor{\bsnm{Wang}, \binits{Z.}},
\bauthor{\bsnm{Ji}, \binits{J.}},
\bauthor{\bsnm{Wang}, \binits{J.}},
\bauthor{\bsnm{Zou}, \binits{W.}},
\bauthor{\bsnm{You}, \binits{D.}},
\bauthor{\bsnm{Qin}, \binits{G.}}:
\batitle{A novel intelligent method based on the gaussian heatmap sampling
  technique and convolutional neural network for landslide susceptibility
  mapping}.
\bjtitle{Remote Sensing}
\bvolume{14},
\bfpage{2866}
(\byear{2022})
\doiurl{10.3390/rs14122866}
\end{barticle}
\endbibitem

\bibitem[\protect\citeauthoryear{Yordanov et~al.}{2021}]{biagi2021}
\begin{barticle}
\bauthor{\bsnm{Yordanov}, \binits{V.}},
\bauthor{\bsnm{Biagi}, \binits{L.}},
\bauthor{\bsnm{Truong}, \binits{X.Q.}},
\bauthor{\bsnm{Tran}, \binits{V.A.}},
\bauthor{\bsnm{Brovelli}, \binits{M.A.}}:
\batitle{An overview of geoinformatics state of the art techniques for
  landslide monitoring and mapping}.
\bjtitle{Int. Arch. Photogramm. Remote Sens. Spatial Inf. Sci.}
\bvolume{XLVI-4/W2-2021},
\bfpage{205}--\blpage{212}
(\byear{2021})
\doiurl{10.5194/isprs-archives-XLVI-4-W2-2021-205-2021}
\end{barticle}
\endbibitem

\bibitem[\protect\citeauthoryear{Chung and Fabbri}{2003}]{validation_paper}
\begin{barticle}
\bauthor{\bsnm{Chung}, \binits{C.-J.}},
\bauthor{\bsnm{Fabbri}, \binits{A.}}:
\batitle{Validation of spatial prediction models for landslide hazard mapping}.
\bjtitle{Natural Hazards}
\bvolume{30},
\bfpage{451}--\blpage{472}
(\byear{2003})
\doiurl{10.1023/B:NHAZ.0000007172.62651.2b}
\end{barticle}
\endbibitem

\bibitem[\protect\citeauthoryear{Muñoz-Torrero~Manchado
  et~al.}{2022}]{land_sus_validated2}
\begin{botherref}
\oauthor{\bsnm{Muñoz-Torrero~Manchado}, \binits{A.}},
\oauthor{\bsnm{Ballesteros-Canovas}, \binits{J.}},
\oauthor{\bsnm{Allen}, \binits{S.}}:
Deforestation controls landslide susceptibility in far-western nepal.
Catena
\textbf{219}
(2022)
\doiurl{10.1016/j.catena.2022.106627}
\end{botherref}
\endbibitem

\bibitem[\protect\citeauthoryear{David et~al.}{1993}]{land_glossary}
\begin{bbook}
\bauthor{\bsnm{David}, \binits{C.}},
\bauthor{\bsnm{Lisandro}, \binits{B.}},
\bauthor{\bsnm{Edmund}, \binits{K.}},
\bauthor{\bsnm{Guy}, \binits{L.}},
\bauthor{\bsnm{G.I.Ter-Stepanian}},
\bauthor{\bsnm{Zhang}, \binits{Z.}}:
\bbtitle{Multilingual Landslide Glossary},
(\byear{1993})
\end{bbook}
\endbibitem

\bibitem[\protect\citeauthoryear{Baddeley}{2006}]{baddeley06}
\begin{botherref}
\oauthor{\bsnm{Baddeley}, \binits{A.}}:
Spatial point processes and their applications.
Stochastic Geometry: Lectures given at the C.I.M.E. 2004, Lecture Notes in
  Mathematics 1892
(2006)
\end{botherref}
\endbibitem

\bibitem[\protect\citeauthoryear{Breiman}{2001}]{RF}
\begin{barticle}
\bauthor{\bsnm{Breiman}, \binits{L.}}:
\batitle{Random forests}.
\bjtitle{Machine Learning}
\bvolume{45},
\bfpage{5}--\blpage{32}
(\byear{2001})
\doiurl{10.1023/A:1010950718922}
\end{barticle}
\endbibitem

\bibitem[\protect\citeauthoryear{Gatti et~al.}{2024}]{GATTI2024112798}
\begin{barticle}
\bauthor{\bsnm{Gatti}, \binits{F.}},
\bauthor{\bsnm{{de Falco}}, \binits{C.}},
\bauthor{\bsnm{Perotto}, \binits{S.}},
\bauthor{\bsnm{Formaggia}, \binits{L.}},
\bauthor{\bsnm{Pastor}, \binits{M.}}:
\batitle{A scalable well-balanced numerical scheme for the modeling of
  two-phase shallow granular landslide consolidation}.
\bjtitle{Journal of Computational Physics}
\bvolume{501},
\bfpage{112798}
(\byear{2024})
\doiurl{10.1016/j.jcp.2024.112798}
\end{barticle}
\endbibitem

\bibitem[\protect\citeauthoryear{Gatti et~al.}{2023}]{GATTI2023105362}
\begin{barticle}
\bauthor{\bsnm{Gatti}, \binits{F.}},
\bauthor{\bsnm{Bonaventura}, \binits{L.}},
\bauthor{\bsnm{Menafoglio}, \binits{A.}},
\bauthor{\bsnm{Papini}, \binits{M.}},
\bauthor{\bsnm{Longoni}, \binits{L.}}:
\batitle{A fully coupled superficial runoff and soil erosion basin scale model
  with efficient time stepping}.
\bjtitle{Computers \& Geosciences}
\bvolume{177},
\bfpage{105362}
(\byear{2023})
\doiurl{10.1016/j.cageo.2023.105362}
\end{barticle}
\endbibitem

\bibitem[\protect\citeauthoryear{Quecedo et~al.}{2004}]{quecedo04}
\begin{barticle}
\bauthor{\bsnm{Quecedo}, \binits{M.}},
\bauthor{\bsnm{Pastor}, \binits{M.}},
\bauthor{\bsnm{Herreros}, \binits{I.}},
\bauthor{\bsnm{Fernandez-Merodo}, \binits{J.A.}}:
\batitle{Numerical modelling of the propagation of fast landslides using the
  finite element method}.
\bjtitle{International Journal for Numerical Methods in Engineering}
\bvolume{59},
\bfpage{755}--\blpage{794}
(\byear{2004})
\doiurl{10.1002/nme.841}
\end{barticle}
\endbibitem

\bibitem[\protect\citeauthoryear{Pastor et~al.}{2017}]{pastor17}
\begin{botherref}
\oauthor{\bsnm{Pastor}, \binits{M.}},
\oauthor{\bsnm{Yague}, \binits{A.}},
\oauthor{\bsnm{Stickle}, \binits{M.}},
\oauthor{\bsnm{Manzanal}, \binits{D.}},
\oauthor{\bsnm{Mira}, \binits{P.}}:
A two-phase sph model for debris flow propagation.
International Journal for Numerical and Analytical Methods in Geomechanics
\textbf{42}
(2017)
\doiurl{10.1002/nag.2748}
\end{botherref}
\endbibitem

\bibitem[\protect\citeauthoryear{Pastor et~al.}{2021}]{pastor21}
\begin{barticle}
\bauthor{\bsnm{Pastor}, \binits{M.}},
\bauthor{\bsnm{Moussavi~Tayyebi}, \binits{S.}},
\bauthor{\bsnm{Stickle}, \binits{M.}},
\bauthor{\bsnm{Yague}, \binits{A.}},
\bauthor{\bsnm{Molinos~Perez}, \binits{M.}},
\bauthor{\bsnm{Navas}, \binits{P.}},
\bauthor{\bsnm{Manzanal}, \binits{D.}}:
\batitle{A depth integrated, coupled, two-phase model for debris flow
  propagation}.
\bjtitle{Acta Geotechnica}
\bvolume{16},
\bfpage{1}--\blpage{25}
(\byear{2021})
\doiurl{10.1007/s11440-020-01114-4}
\end{barticle}
\endbibitem

\bibitem[\protect\citeauthoryear{Gatti et~al.}{2024}]{GATTI2024128525}
\begin{barticle}
\bauthor{\bsnm{Gatti}, \binits{F.}},
\bauthor{\bsnm{{de Falco}}, \binits{C.}},
\bauthor{\bsnm{Perotto}, \binits{S.}},
\bauthor{\bsnm{Formaggia}, \binits{L.}}:
\batitle{A scalable well-balanced numerical scheme for the simulation of fast
  landslides with efficient time stepping}.
\bjtitle{Applied Mathematics and Computation}
\bvolume{468},
\bfpage{128525}
(\byear{2024})
\doiurl{10.1016/j.amc.2023.128525}
\end{barticle}
\endbibitem

\bibitem[\protect\citeauthoryear{Ballerine}{2017}]{twi_report}
\begin{botherref}
\oauthor{\bsnm{Ballerine}, \binits{C.}}:
Topographic Wetness Index Urban Flooding Awareness Act Action Support
\end{botherref}
\endbibitem

\bibitem[\protect\citeauthoryear{John~Weier}{2000}]{ndvi_report}
\begin{botherref}
\oauthor{\bsnm{John~Weier}, \binits{D.H.}}:
Measuring Vegetation (NDVI and EVI).
\url{https://earthobservatory.nasa.gov/features/MeasuringVegetation}.
[Online; accessed 09-03-2023]
(2000)
\end{botherref}
\endbibitem

\bibitem[\protect\citeauthoryear{Royden and Fitzpatrick}{2018 - 2010}]{radnik}
\begin{bbook}
\bauthor{\bsnm{Royden}, \binits{H.L.}},
\bauthor{\bsnm{Fitzpatrick}, \binits{P.}}:
\bbtitle{Real Analysis / H.L. Royden, Stanford University, P.M. Fitzpatrick,
  University of Maryland, College Park.},
\bedition{Fourth edition [2018 reissue].} edn.
\bsertitle{Pearson modern classic}.
\bpublisher{Pearson},
\blocation{New York, NY}
(\byear{2018 - 2010})
\end{bbook}
\endbibitem

\bibitem[\protect\citeauthoryear{Coeurjolly and
  Lavancier}{2019}]{coeurjoully19}
\begin{bbook}
\bauthor{\bsnm{Coeurjolly}, \binits{J.-F.}},
\bauthor{\bsnm{Lavancier}, \binits{F.}}:
\bbtitle{Understanding Spatial Point Patterns Through Intensity and Conditional
  Intensities},
pp. \bfpage{45}--\blpage{85}
(\byear{2019}).
\doiurl{10.1007/978-3-030-13547-8_2}
\end{bbook}
\endbibitem

\bibitem[\protect\citeauthoryear{Yue}{2015}]{yue_yu15}
\begin{barticle}
\bauthor{\bsnm{Yue}, \binits{Y.}}:
\batitle{Variable selection for inhomogeneous spatial point process models}.
\bjtitle{Canadian Journal of Statistics}
\bvolume{43},
\bfpage{288}--\blpage{305}
(\byear{2015})
\end{barticle}
\endbibitem

\bibitem[\protect\citeauthoryear{Hastie and Tibshirani}{1986}]{hastie_tib_GAM}
\begin{barticle}
\bauthor{\bsnm{Hastie}, \binits{T.}},
\bauthor{\bsnm{Tibshirani}, \binits{R.}}:
\batitle{Generalized additive models}.
\bjtitle{Statistical Science}
\bvolume{1}(\bissue{3}),
\bfpage{297}--\blpage{310}
(\byear{1986}).
Accessed 2023-03-13
\end{barticle}
\endbibitem

\bibitem[\protect\citeauthoryear{Wood}{2017}]{wood17}
\begin{bbook}
\bauthor{\bsnm{Wood}, \binits{S.N.}}:
\bbtitle{Generalized Additive Models. An Introduction with R}.
\bpublisher{Chapman and Hall/CRC},
\blocation{New York}
(\byear{2017})
\end{bbook}
\endbibitem

\bibitem[\protect\citeauthoryear{Cowling et~al.}{1997}]{bootPois}
\begin{botherref}
\oauthor{\bsnm{Cowling}, \binits{A.}},
\oauthor{\bsnm{Hall}, \binits{P.}},
\oauthor{\bsnm{Phillips}, \binits{M.}}:
Bootstrap confidence regions for the intensity of a poisson point process.
Journal of the American Statistical Association
\textbf{91}
(1997)
\doiurl{10.1080/01621459.1996.10476719}
\end{botherref}
\endbibitem

\bibitem[\protect\citeauthoryear{Borrajo et~al.}{2018}]{bootCov}
\begin{botherref}
\oauthor{\bsnm{Borrajo}, \binits{M.}},
\oauthor{\bsnm{Gonzãlez-Manteiga}, \binits{W.}},
\oauthor{\bsnm{Miranda}, \binits{M.}}:
Bootstrapping kernel intensity estimation for nonhomogeneous point processes
  depending on spatial covariates.
Computational Statistics \& Data Analysis
\textbf{144}
(2018)
\doiurl{10.1016/j.csda.2019.106875}
\end{botherref}
\endbibitem

\bibitem[\protect\citeauthoryear{Turner and Baddeley}{2005}]{spatstat}
\begin{botherref}
\oauthor{\bsnm{Turner}, \binits{R.}},
\oauthor{\bsnm{Baddeley}, \binits{A.}}:
Spatstat: an r package for analyzing spatial point patterns.
Journal of Statistical Software
\textbf{12}
(2005)
\doiurl{10.18637/jss.v012.i06}
\end{botherref}
\endbibitem

\bibitem[\protect\citeauthoryear{Basu et~al.}{2017}]{iRF}
\begin{botherref}
\oauthor{\bsnm{Basu}, \binits{S.}},
\oauthor{\bsnm{Kumbier}, \binits{K.}},
\oauthor{\bsnm{Brown}, \binits{J.}},
\oauthor{\bsnm{Yu}, \binits{B.}}:
Iterative random forests to detect predictive and stable high-order
  interactions.
Proceedings of the National Academy of Sciences
\textbf{115}
(2017)
\doiurl{10.1073/pnas.1711236115}
\end{botherref}
\endbibitem

\bibitem[\protect\citeauthoryear{Di~Napoli et~al.}{2021}]{shallow_slope}
\begin{botherref}
\oauthor{\bsnm{Di~Napoli}, \binits{M.}},
\oauthor{\bsnm{Di~Martire}, \binits{D.}},
\oauthor{\bsnm{Bausilio}, \binits{G.}},
\oauthor{\bsnm{Calcaterra}, \binits{D.}},
\oauthor{\bsnm{Confuorto}, \binits{P.}},
\oauthor{\bsnm{Firpo}, \binits{M.}},
\oauthor{\bsnm{Pepe}, \binits{G.}},
\oauthor{\bsnm{Cevasco}, \binits{A.}}:
Rainfall-induced shallow landslide detachment, transit and runout
  susceptibility mapping by integrating machine learning techniques and
  gis-based approaches.
Water
\textbf{13}(4)
(2021)
\doiurl{10.3390/w13040488}
\end{botherref}
\endbibitem

\bibitem[\protect\citeauthoryear{Çellek}{2022}]{slope_land}
\begin{barticle}
\bauthor{\bsnm{Çellek}, \binits{S.}}:
\batitle{Effect of the slope angle and its classification on landslides}.
\bjtitle{Himalayan Geology}
\bvolume{43},
\bfpage{85}--\blpage{95}
(\byear{2022})
\end{barticle}
\endbibitem

\bibitem[\protect\citeauthoryear{Rickli and Graf}{2009}]{switz_shallow}
\begin{botherref}
\oauthor{\bsnm{Rickli}, \binits{C.}},
\oauthor{\bsnm{Graf}, \binits{F.}}:
Effects of forests on shallow landslides - case studies in switzerland.
For. Snow Landscape Res.
\textbf{82}
(2009)
\end{botherref}
\endbibitem

\bibitem[\protect\citeauthoryear{Lee and Min}{2001}]{suscept_korea}
\begin{barticle}
\bauthor{\bsnm{Lee}, \binits{S.}},
\bauthor{\bsnm{Min}, \binits{K.}}:
\batitle{Statistical analysis of landslide susceptibility at yongin, korea}.
\bjtitle{Environmental Geology}
\bvolume{40},
\bfpage{1095}--\blpage{1113}
(\byear{2001})
\doiurl{10.1007/s002540100310}
\end{barticle}
\endbibitem

\bibitem[\protect\citeauthoryear{Grabowski et~al.}{2022}]{suscept_poland}
\begin{barticle}
\bauthor{\bsnm{Grabowski}, \binits{D.}},
\bauthor{\bsnm{Laskowicz}, \binits{I.}},
\bauthor{\bsnm{Malka}, \binits{A.}},
\bauthor{\bsnm{Rubinkiewicz}, \binits{J.}}:
\batitle{Geoenvironmental conditioning of landsliding in river valleys of
  lowland regions and its significance in landslide susceptibility assessment:
  A case study in the lower vistula valley, northern poland}.
\bjtitle{Geomorphology}
\bvolume{419},
\bfpage{108490}
(\byear{2022})
\doiurl{10.1016/j.geomorph.2022.108490}
\end{barticle}
\endbibitem

\bibitem[\protect\citeauthoryear{Today}{2019}]{article_sondrio}
\begin{botherref}
\oauthor{\bsnm{Today}, \binits{S.}}:
After 32 years Valtellina do not forget the flood in 1987.
\url{https://www.sondriotoday.it/cronaca/alluvione-valtellina-1987.html}.
[Online; accessed 22-03-2023]
(2019)
\end{botherref}
\endbibitem

\bibitem[\protect\citeauthoryear{González et~al.}{2016}]{spatiotemp}
\begin{barticle}
\bauthor{\bsnm{González}, \binits{J.A.}},
\bauthor{\bsnm{Rodríguez-Cortés}, \binits{F.J.}},
\bauthor{\bsnm{Cronie}, \binits{O.}},
\bauthor{\bsnm{Mateu}, \binits{J.}}:
\batitle{Spatio-temporal point process statistics: A review}.
\bjtitle{Spatial Statistics}
\bvolume{18},
\bfpage{505}--\blpage{544}
(\byear{2016})
\doiurl{10.1016/j.spasta.2016.10.002}
\end{barticle}
\endbibitem

\bibitem[\protect\citeauthoryear{Baddeley et~al.}{2005}]{baddeley05}
\begin{barticle}
\bauthor{\bsnm{Baddeley}, \binits{A.J.}},
\bauthor{\bsnm{Turner}, \binits{R.}},
\bauthor{\bsnm{Møller}, \binits{J.}},
\bauthor{\bsnm{Hazelton}, \binits{M.}}:
\batitle{Residual analysis for spatial point processes (with discussion)}.
\bjtitle{Journal of the Royal Statistical Society: Series B (Statistical
  Methodology)}
\bvolume{67},
\bfpage{617}--\blpage{666}
(\byear{2005})
\doiurl{10.1111/j.1467-9868.2005.00519.x}
\end{barticle}
\endbibitem

\bibitem[\protect\citeauthoryear{Baddeley}{2007}]{baddeley07}
\begin{botherref}
\oauthor{\bsnm{Baddeley}, \binits{A.}}:
Validation of statistical models for spatial point patterns
\textbf{371},
22
(2007)
\end{botherref}
\endbibitem

\end{thebibliography}

\end{document}